\documentclass[journal]{IEEEtran}

\usepackage[T1]{fontenc}
\usepackage{graphicx}
\usepackage{cite}
\usepackage{subcaption}

\usepackage{overpic}
\usepackage{amssymb}
\usepackage{amsmath}
\usepackage{array}
\usepackage{color}
\usepackage{acronym}

\IEEEoverridecommandlockouts

\newtheorem{definition}{Definition}

\acrodef{snr}[SNR]{signal-to-noise ratio}
\acrodef{gnss}[GNSS]{global navigation satellite system}
\acrodef{los}[LoS]{line-of-sight}
\acrodef{aoa}[AoA]{angle-of-arrival}
\acrodef{aod}[AoD]{angle-of-departure}
\acrodef{uav}[UAV]{unmanned aerial vehicle}
\acrodef{uavs}[UAVs]{unmanned aerial vehicles}
\acrodef{d2d}[D2D]{device-to-device}
\acrodef{slam}[SLAM]{Simultaneous localization and mapping}

\begin{document}

\title{Massive MIMO is a Reality---What is Next? \\[+1mm] \huge \emph{ Five Promising Research Directions for Antenna Arrays}}

\author{ \vspace{+4mm}
\IEEEauthorblockN{Emil Bj\"ornson, Luca Sanguinetti, Henk Wymeersch, Jakob Hoydis, Thomas L. Marzetta
\thanks{
\newline \indent E.~Bj\"ornson is with the Department of Electrical Engineering (ISY), Link\"{o}ping University, 58183 Link\"{o}ping, Sweden (emil.bjornson@liu.se). He was supported by ELLIIT and CENIIT.
\newline \indent L.~Sanguinetti is with the University of Pisa, Dipartimento di Ingegneria dell'Informazione, 56122 Pisa, Italy (luca.sanguinetti@unipi.it). 
\newline \indent H.~Wymeersch is with the Department of Electrical Engineering, Chalmers University of Technology, 412 96 Gothenburg, Sweden (henkw@chalmers.se). 
\newline \indent
J.~Hoydis is with Nokia Bell Labs, Paris-Saclay, 91620 Nozay, France (jakob.hoydis@nokia-bell-labs.com).
\newline \indent T.~L.~Marzetta is with the Department of Electrical and Computer Engineering, New York University, Tandon School of Engineering, Brooklyn, NY (tom.marzetta@nyu.edu).}
}}

\maketitle

\begin{abstract}
Massive MIMO (multiple-input multiple-output) is no longer a ``wild'' or ``promising''  concept for future cellular networks---in 2018 it became a reality. Base stations (BSs) with 64 fully digital transceiver chains were commercially deployed in several countries, the key ingredients of Massive MIMO have made it into the 5G standard, the signal processing methods required to achieve unprecedented spectral efficiency have been developed, and the limitation due to pilot contamination has been resolved. Even the development of fully digital Massive MIMO arrays for mmWave frequencies---once viewed prohibitively complicated and costly---is well underway. In a few years, Massive MIMO with fully digital transceivers will be a mainstream feature at both sub-6\,GHz and mmWave frequencies.

In this paper, we explain how the first chapter of the Massive MIMO research saga has come to an end, while the story has just begun. The coming wide-scale deployment of BSs with massive antenna arrays opens the door to a brand new world where spatial processing capabilities are omnipresent. In addition to mobile broadband services, the antennas can be used for other communication applications, such as low-power machine-type or ultra-reliable communications, as well as non-communication applications such as radar, sensing and positioning. We outline five new Massive MIMO related research directions: Extremely large aperture arrays, Holographic Massive MIMO, Six-dimensional positioning, Large-scale MIMO radar, and Intelligent Massive MIMO.
\end{abstract}


\IEEEpeerreviewmaketitle

\section{Introduction}

\begin{figure}[t!]
	\centering \vspace{-2mm}
	\begin{overpic}[width=.8\columnwidth,tics=10]{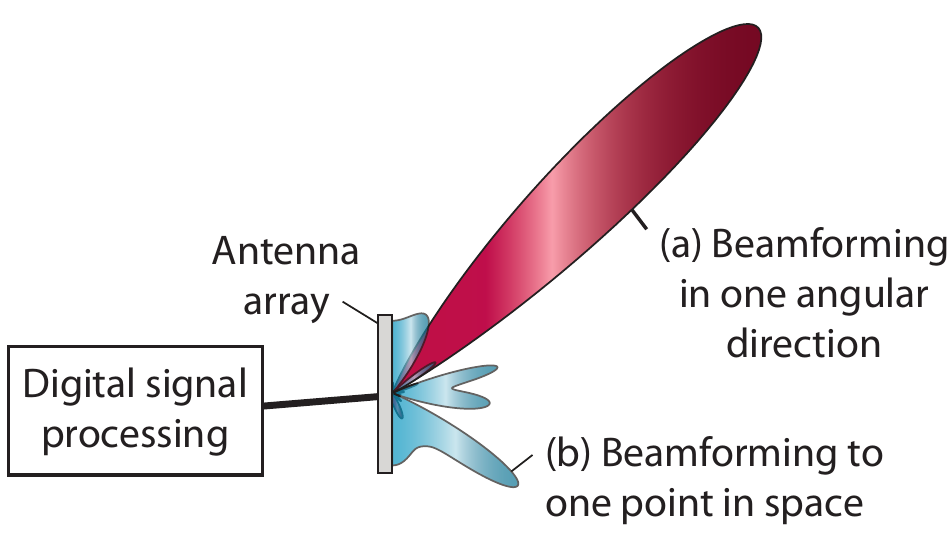}
\end{overpic} 
	\caption{Beamforming from an antenna array can be used to (a) focus the radiated signal in one angular direction or (b) focus the signal at one particular point in space, in which case the radiated signal might have no dominant directivity.
 The radiation patterns in this figure were computed using eight-antenna uniform linear arrays.}
	\label{fig:beamforming}
\end{figure}

While an individual antenna has a fixed radiation pattern, antenna arrays are capable of changing their radiation patterns over time and frequency, for both transmission and reception. This is traditionally illustrated as the formation of spatial beams in one (or a few) distinct angular directions, as shown in Fig.~\ref{fig:beamforming}(a), but antenna arrays are also capable of many other types of spatial filtering. For example, the signal processing that controls the array can be used to focus a signal at an arbitrary point in space which, in a rich multi-path propagation environment, corresponds to emitting a superposition of many angular beams so that the radiated pattern has no dominant directivity, as shown in Fig.~\ref{fig:beamforming}(b). Both examples are commonly referred to as \emph{beamforming}, even if a ``beam'' is strictly speaking only created in the former case. In addition, the array can be used to sense the propagation environment, for example, to detect anomalies or moving objects.

Many different applications for antenna/sensor arrays have been conceived over the years. The 1988 overview paper \cite{Veen1998a} by Van Veen and Buckley outlined radar, sonar, communications, imaging, geophysical exploration, astrophysical exploration, and biomedical applications that were identified in the 70s and 80s. The signal processing methods that are nowadays known as maximum ratio (MR), zero-forcing (ZF), and minimum-mean square error (MMSE) processing were already known at that time, but under different names. When writing this article---30 years later---the recent textbooks \cite{Marzetta2016a,massivemimobook,heath_lozano_2018} on multiple antenna communications are still treating MR, ZF, and MMSE as the state-of-the-art methods. With that in mind, one might wonder: what has the research community been doing the past 30 years?

The devil is in the details. Every application has its unique characteristics and it is hard to bridge the divide between an initial concept and a successful commercial solution. Let us take a closer look at the development of multi-user MIMO, by which we refer to communication systems that use antenna arrays at the BSs to spatially multiplex several users at the same time-frequency resource. In a paper from 1987 \cite{Winters1987a}, Winters described that one can use antenna arrays to discriminate between uplink signals from different users by spatial processing, called \emph{receive combining}. A few years later \cite{Swales1990a}, Swales et al.~described how antenna arrays can be also used to spatially multiplex users in the downlink, in which case the spatial processing is called \emph{transmit precoding}.

\begin{figure}[t!]
	\centering \vspace{-2mm}
	\begin{overpic}[width=.8\columnwidth,tics=10]{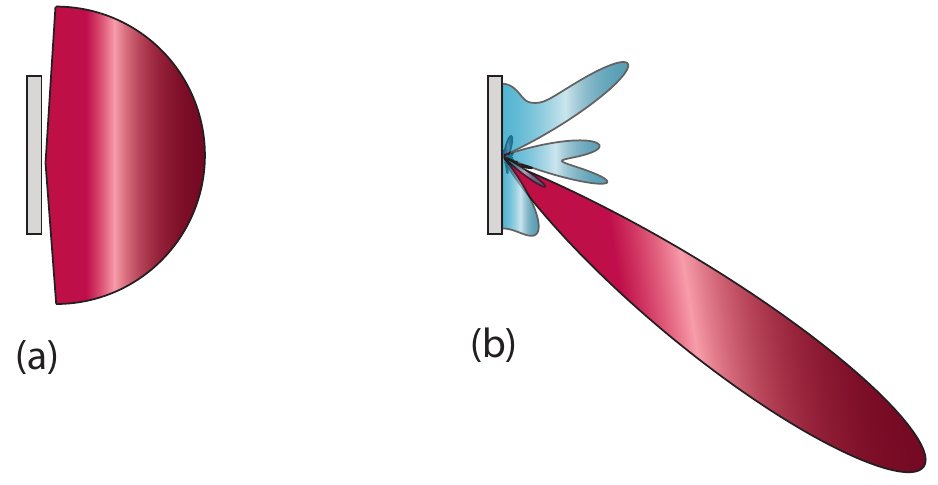}
\end{overpic} 
	\caption{A classical BS radiates one signal uniformly into its coverage area, as shown in (a). A BS capable of multi-user MIMO can radiate multiple signals (indicated by different colors) that are focused at their respective receivers, as shown in (b).}
	\label{fig:mu-mimo}
\end{figure}

The difference between classical communication systems and multi-user MIMO  can be seen by comparing Fig.~\ref{fig:mu-mimo}(a) and Fig.~\ref{fig:mu-mimo}(b). Instead of radiating one signal ``uniformly'' into the coverage area of the transmitter, an antenna array can focus the same signal at its intended receiver. If $M$ antennas are used, an $M$ times stronger signal can (ideally) be achieved at the receiver, without changing the radiated power. The consequence is that less signal power is observed at other places and one can, therefore, focus other signals towards other points in space without causing much interference between transmissions. If $K$ users are spatially multiplexed in the downlink, each user might be allocated only $1/K$ of the total radiated power, but if $M \geq K$, the beamforming will still make the received signal $M/K>1$ times stronger than in the classical system. Hence, the overall spectral efficiency [b/s/Hz] of the system grows proportionally to the number of users if $M$ and $K$ are jointly increased \cite{massivemimobook}. That is why $M \gg K$ is the preferable operating regime for multi-user MIMO. Note that the word ``multiple'' in the MIMO acronym refers to the multiple antennas at the BS and the multiple users, while each user can have any number of antennas.

Extensive multi-user MIMO field trials were carried out in the 90s \cite{Anderson1999a} and ArrayComm deployed commercial products in Japan \cite[Example 10.1]{Tse2005a}. However, the multi-user MIMO technology (then known as spatial division multiple access (SDMA) \cite{Richard1996a}) did not become a commercial success in the 90s or 00s. There are many contributing factors and their relative impact is debatable, but we will mention a few key factors:

\begin{enumerate}

\item In a time when circuit-switched low-rate voice communication was the dominant service, it was not the spectral efficiency but the capability of multiplexing a certain number of users per km$^2$ that was the key performance metric. Since multi-user MIMO was rather complicated and expensive to implement in the 90s, it was simply more cost-efficient to deploy more BSs using classical hardware than to invest in the new and rather untested multi-user MIMO technology. Since classical BSs rely on orthogonal time-frequency scheduling, each BS could multiplex a much smaller number of users than a BS supporting multi-user MIMO. Nevertheless, a denser deployment can make the total number of multiplexed users per km$^2$ the same as with a less dense multi-user MIMO deployment.

\item The technological development was largely driven by heuristics and experiments since the first major information theoretic breakthroughs for multi-user MIMO were made in the 00s \cite{Caire2003a,Goldsmith2003a,Viswanath2003a}. 
The early papers considered perfect channel state information (CSI) and it took many more years for {the information theory literature} to provide guidance on how to deal with the imperfect CSI that occurs in any practical communication system.

\item Most telecom operators had frequency-division duplex (FDD) licenses at the time and it is hard to acquire downlink CSI that is sufficiently accurate for beamforming in such systems \cite{Bjornson2016b}. Beamforming based on \ac{aoa} estimation and quantization codebooks were considered to send  beams in roughly the ``right'' way. This worked rather well for single-user systems, but the accuracy was insufficient to control inter-user interference.

\item The number of antennas was fairly small ($M\approx 8$) which is insufficient to achieve the spatial resolution that is necessary for effective interference suppression, using ZF or MMSE processing, even with perfect CSI. 

\end{enumerate}

These factors created a negative attitude against the multi-user MIMO technology, which has partially remained until today, even if the technology has changed dramatically during the last decade.

\subsection{Massive MIMO is a Reality}

To address the shortcomings of conventional multi-user MIMO, the Massive MIMO concept was introduced in \cite{Marzetta2010a}. It is now well accepted that many of the previous challenges can be resolved by equipping the BSs with very large numbers of antennas and utilizing time-division duplex (TDD) operation and {the} uplink-downlink channel reciprocity to achieve a communication protocol that supports arbitrary antenna numbers and channel conditions:
\begin{itemize}
\item Each antenna can be built using inexpensive handset-grade hardware components \cite{Bjornson2014a}, which keeps the cost down.

\item The communication design is deeply rooted in information theory and finds the right operating regime by taking the properties of imperfect CSI  into account \cite{Marzetta2016a,massivemimobook}.

\item The signal processing complexity is manageable if dedicated circuits are designed \cite{Prabhu2017a,Perre2018a}.

\item There is no need to rely on \ac{aoa} estimation or quantization codebooks, which only work well for channels with angular sparsity and calibrated array structures.

\item A large number of antennas ($M\geq 64$) leads to an unprecedented spatial resolution, robustness against small-scale fading, and the ability to spatially suppress interference even with imperfect CSI if  $M\gg K$.

\end{itemize}

A solid theory for {Massive MIMO in block-fading channels} has been developed in recent years, thanks to the contributions of many researchers in academia and industry. Some of the key research directions that are now approaching the finish lines are the spectral efficiency analysis \cite{Ngo2013a,Yang2013c,Hoydis2013a,Bjornson2016a}, system design for high energy efficiency \cite{Yang2013a,Mohammed2014a,Bjornson2015a}, {pilot contamination and decontamination \cite{Jose2011b,Mueller2014b,Yin2016a,BHS18A,Neumann2018a},} and power optimization \cite{Guo2014a,Saxena2015a,Cheng2017a}. 
The maturity of the research on Massive MIMO is underlined by the two recent textbooks on the topic that cover the fundamentals  \cite{Marzetta2016a} as well as advanced topics \cite{massivemimobook}. {The research in the aforementioned directions can certainly continue under more practical modeling assumptions, but the main point is that the basics are well understood and noncontroversial.} 

Nevertheless, Massive MIMO has been (and still is) met with skepticism and many misconceptions have flourished \cite{Bjornson2016b}, often rooted in the negative past experiences with multi-user MIMO. For example, Massive MIMO has been accused of being too complicated and expensive to implement, as if the transceiver hardware technology had not evolved since the 90s. This belief also spurred large investments into ``less expensive'' hybrid analog-digital array implementations and complicated beam-searching and beam-tracking algorithms that are needed to operate such arrays. The premise for this development is the belief that fully digital transceiver chains are practically infeasible to implement---particularly in mmWave bands but also at sub-6\,GHz bands---but is this correct?

\begin{figure}[t!]
	\centering \vspace{-2mm}
	\begin{overpic}[width=\columnwidth,tics=10]{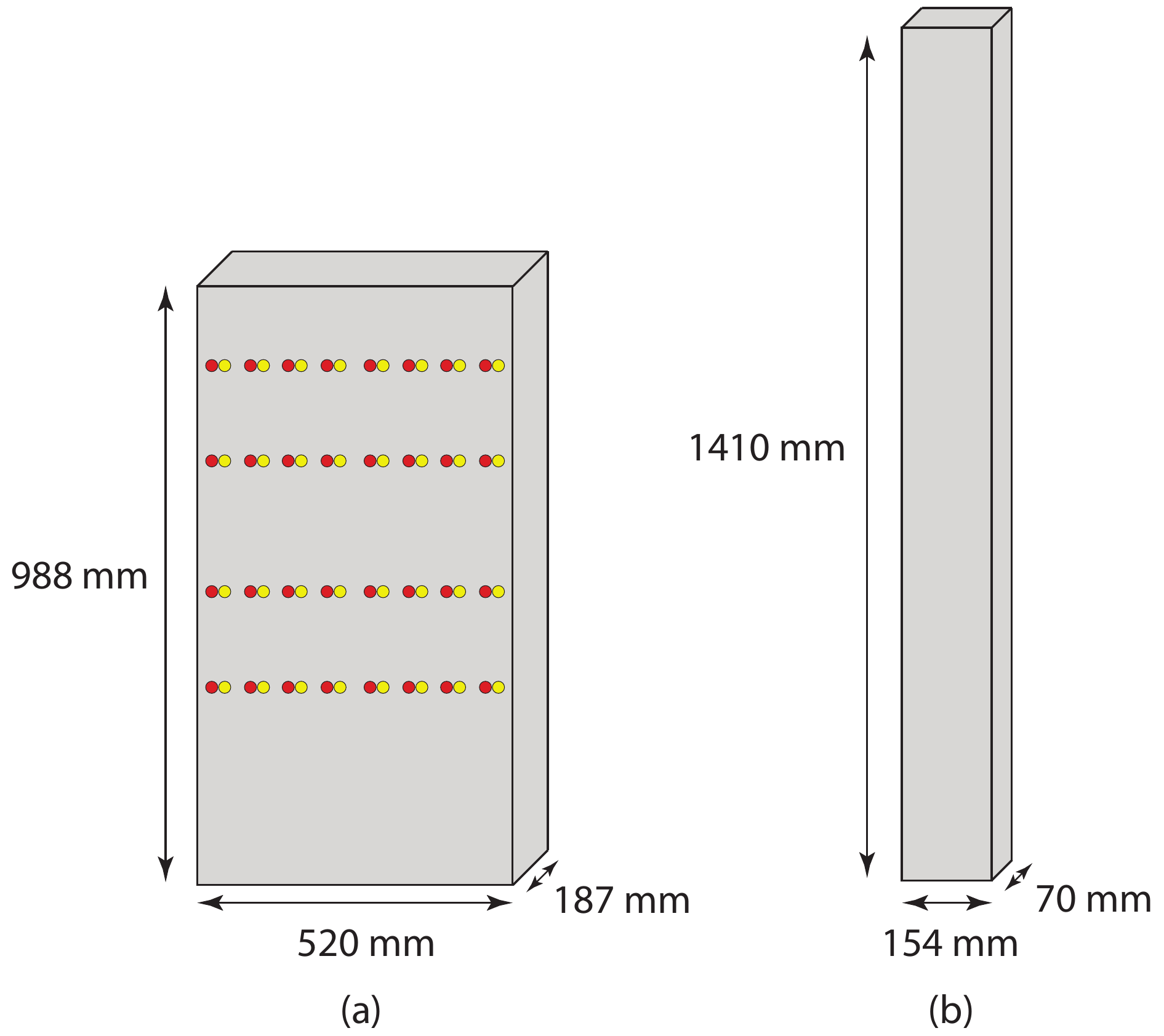}
\end{overpic} 
	\caption{Comparison of the form factors of (a) the Ericsson AIR 6468, 64-antenna array; (b) the Kathrein 80010621 antenna panel with 16 dBi directivity. Both arrays support the 2.5\,GHz band. The 64 antennas in (a) have varying polarization, indicated by two colors. There is no massive difference in size since 64 low-gain antennas in (a) are compared with one high-gain antenna in (b).}
	\label{fig:AIR6468}
\end{figure}

In 2014, the real-time testbed at Lund University showed that Massive MIMO with 100 fully digital transceiver chains can be implemented using off-the-shelf hardware, requiring ``only'' a major engineering effort \cite{Vieira2014a}. In 2018, the FCC approved the first line of Massive MIMO products, including the Ericsson AIR 6468 \cite{EricssonAIR6468}. This product has 64 antennas connected to 64 fully digital transceiver chains in both uplink and downlink, and it is designed for 4G LTE, so it is even a pre-5G product. The AIR 6468 can be used in either the 2.5\,GHz or 3.5\,GHz band. Compared to a conventional fixed-beam sector antenna designed for the same bands, the Massive MIMO array is wider but has a smaller height, as illustrated in Fig.~\ref{fig:AIR6468}. There is no massive difference in size since the many low-gain antennas in Massive MIMO must be compared with one high-gain antenna. 
The 64 antennas are deployed on 4 rows, each containing 8 dual-polarized antennas \cite{Butovitsch2018a}. This two-dimensional configuration also makes the array compact compared to the large one-dimensional uniform linear arrays that are commonly considered in the academic literature.

The AIR 6468 can aggregate up to three carriers (15-20\,MHz each), supports reciprocity-based beamforming, and signaling using QPSK, 16-QAM, 64-QAM, and 256-QAM. The maximum radiated power is 1.875\,W per antenna, which corresponds to 120 W in total for the array.

The Ericsson AIR 6468 is not unique: Huawei AAU and Nokia Airscale are two competing product lines. Many telecom operators started to deploy this type of array in 2018, including the US operator Sprint that even used the Massive MIMO term in its marketing towards the end users \cite{Sprint2018feb}. {Huawei reported at the Mobile World Congress 2019 that 95\% of their current commercial shipments has either 32 or 64 antennas \cite{Huawei95percent}.} This demonstrates that Massive MIMO is now a reality for cellular networks operating in conventional sub-6\,GHz bands. Hence, the previous claim of Massive MIMO being ``too complicated and expensive to implement'' has finally been disproved.

Since there is no need to standardize which signal processing methods will be used for beamforming and channel estimation at the BS, the solutions implemented in real networks can change over time. The first Massive MIMO products are (probably) using fairly simple signal processing methods, such as MR for beamforming and least-squares for channel estimation. The reason is that most current cells do not have enough simultaneously active users to benefit much from spatial multiplexing. Instead, the telecom operators are mainly observing a need for improving the performance at the cell edge, so basic beamforming to arbitrary points in space, as illustrated in Fig.~\ref{fig:beamforming}(b), is the feature that is implemented first. However, when the antenna arrays have been deployed, the spectral efficiency can be improved by a software update that switches to more advanced methods from the Massive MIMO literature \cite{massivemimobook}. 
A gradual refinement makes practical sense: less complicated methods are easier to implement, more advanced methods are not needed until the number of simultaneously active users and their traffic demands surpass the limits of the simpler methods, and then the more advanced methods can be sold as feature upgrades. {Note that 3G and 4G networks have improved their spectral efficiency in similar ways.}

At mmWave frequencies, the first experimental verification of fully digital antenna arrays was presented in 2018. NEC has developed a 24-antenna uniform linear array that supports digital beamforming in the 28\,GHz band \cite{Tawa2018a}. This prototype will likely be turned into a commercial product in the next few years. {Hence, even if the first 5G products for mmWave communications rely on analog or hybrid implementations to quickly reach the market, it is only a matter of time before digital solutions prevail and become the most cost and energy efficient implementations, thanks to the fast development in semiconductor technology.} This should come as no surprise---in the time when digitalization is embraced in every part of the society, why would cellular technology suddenly go in the opposite direction?

\subsection{What is Next?}

The development of Massive MIMO communication technology is now in the hands of the product departments of companies such as Ericsson, Huawei, Nokia, etc. A large number of communication, signal processing, and optimization algorithms have been developed over the years and it remains to be seen which ones will work well in practice. Modeling simplifications that have been made in academia (e.g., block-fading channels with stochastic small-scale fading or deterministic channel models with angular sparsity) might prevent a straightforward transfer from theory to practical implementation. {Before the product developers have had the chance to try out the existing algorithms, it is hard to tell what further algorithmic development is actually needed. The MIMO research community should certainly support the product developers in their efforts to implement existing algorithms under practical, hardware-related and regulatory constraints. At the same time, it is important to initiate more forward-looking research that considers new applications of antenna arrays that might become the foundation for beyond 5G networks.}

If 5G becomes a commercial success, massive digitally controllable antenna arrays will be deployed ``everywhere''. Conventional sites operating in the sub-6\,GHz band will be equipped with arrays of 64 or more antennas (per sector) to provide spatial multiplexing over wide areas, while new BSs operating in the mmWave bands will be deployed indoors and at the street level to provide local area coverage. The network equipment that controls these antennas has access to an unprecedented spatial resolution in terms of emitting waveforms that give constructive interference at particular points in space and resolving the fine details of received waveforms. What else can we use this spatial resolution for, beyond mobile broadband applications, and how will the antenna {deployments} evolve in the future?

The remainder of this article will consider five forward-looking research directions that aim at using antenna arrays for new non-communication applications and deployment concepts that open up new exciting  possibilities but also pose interesting research challenges.

\section{Direction 1: Extremely Large Aperture Arrays}

\begin{figure}[t!]
	\centering \vspace{-2mm}
	\begin{overpic}[width=\columnwidth,tics=10]{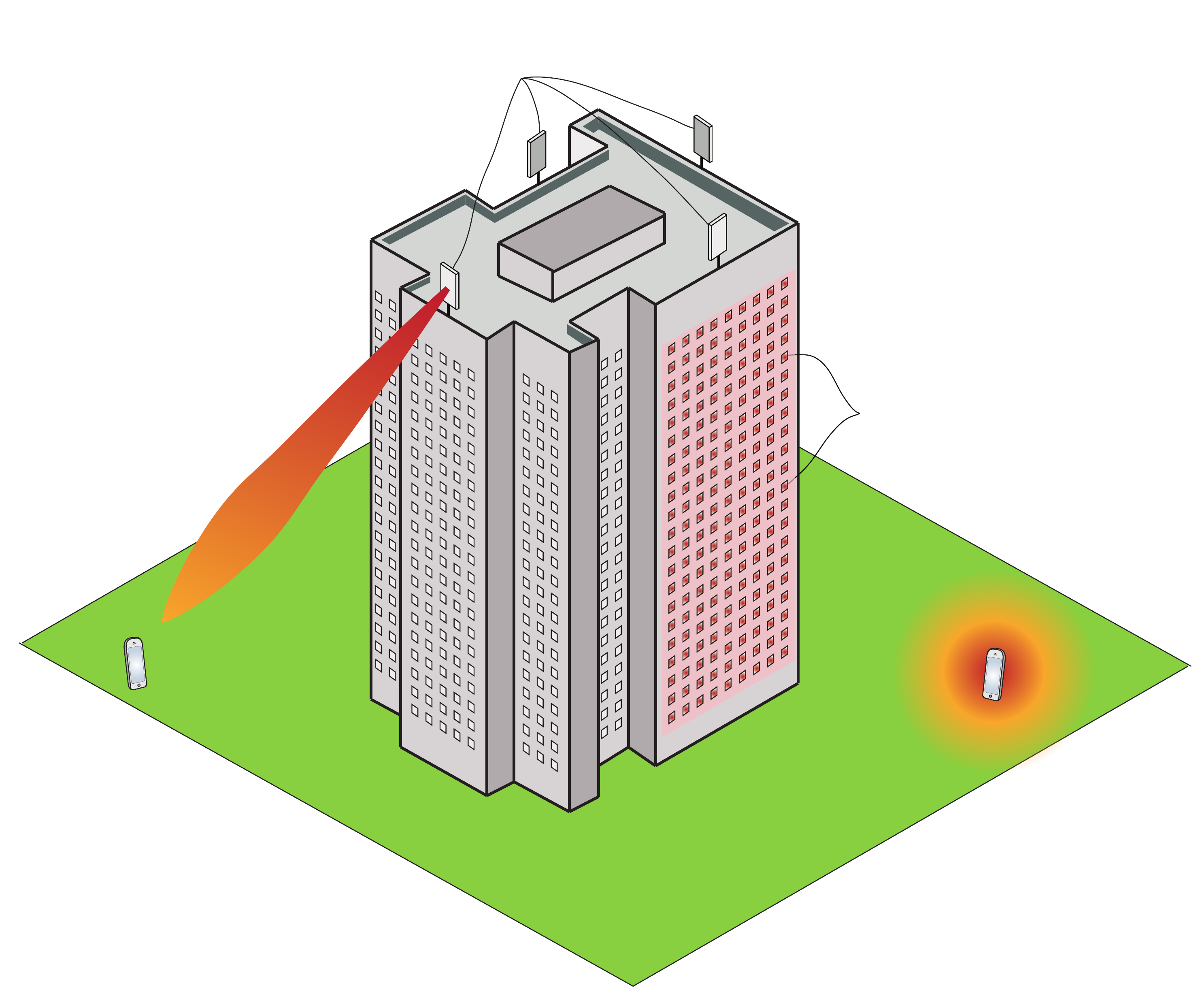}
		 \put (7,78) {(a) Compact co-located Massive MIMO arrays}
	  \put (71,49.5) {(b) Extremely large}
	  \put (77,45.5) {aperture array}
\end{overpic} 
	\caption{The first deployments of Massive MIMO use compact arrays, as shown in (a). The users are in the far-field of the array and, thus, the transmission to a \ac{los} user takes the form of a signal beam. To deploy arrays with very many antennas, we can instead create ELAAs where the antennas are distributed over a large area and hidden into other construction elements, for example, windows as in (b). The user might be in the near-field of the array and then LoS users will not observe signal beams.}
	\label{fig:house}
\end{figure}

The antenna separation in an array is of the order of the wavelength $\lambda$ and the users are located in the far-field of the BS array. These are two classical assumptions in the array processing and wireless communication literature that used to be reasonable but need to be revised going forward. The spectral efficiency of Massive MIMO grows monotonically with the number of antennas \cite{BHS18A}. Thus, we can expect a future where hundreds or thousands of antennas are used to serve a set of users. There are, however, practical limits to how many antennas can be deployed at conventional towers and rooftop locations, for example, determined by the array dimensions allowed by the site owner, the weight, and the wind load. The rather compact 64-antenna product shown in Fig.~\ref{fig:AIR6468}(a) has already been deployed and we will likely see deployments of somewhat larger arrays at some locations as well. Nevertheless, the spatial multiplexing capability of these two-dimensional planar arrays in our three-dimensional world is far from what has been demonstrated in the academic literature, where large one-dimensional arrays are often considered in a two-dimensional world.
In many practical deployment scenarios, the user channels are mainly separable in the horizontal domain \cite{Butovitsch2018a} since the variations in elevation angle between different users and scattering objects are relatively small. The existing 64-antenna products have only eight antennas per horizontal row---how massive is that? Since multi-sector sites are the norm in cellular networks, co-location of three or more compact arrays that point in different directions are also likely to happen, as illustrated in Fig.~\ref{fig:house}(a). However, to deploy more than a few hundred antennas per site and to obtain a truly massive spatial resolution in the horizontal domain, we need new antenna deployment strategies.

Instead of gathering all the antennas in a single box, which will be visible and heavy, the antennas can be distributed over a substantially larger area and made invisible by integrating them into existing construction elements. Fig.~\ref{fig:house}(b) exemplifies a setup where the antennas are deployed next to each window in a tall building. If one dual-polarized antenna is hidden in each corner of the window, there are 1512 antennas in this example. Suppose the adjacent windows are 3\,m apart, then the array spans an area of $24$\,m $\times$ $60$\,m. This is an \emph{Extremely Large Aperture Array} (ELAA) \cite{Amiri2018a} compared to the conventional compact arrays illustrated in Fig.~\ref{fig:AIR6468}; the antenna separation is at the order of meters, which is much larger than the wavelength (being in the range from one decimeter down to a few millimeters in the frequency ranges considered in 5G). Another example of an ELAA is when the antennas are distributed over a large geographical area so that each user is essentially surrounded by BS antennas, rather than the conventional case of each BS being surrounded by users. This has recently been called \emph{Cell-free Massive MIMO} \cite{Ngo2017b,Nayebi2017a,Bjornson2019c}, but the basic concept has its roots in papers on {distributed MIMO} from the early 00s \cite{Shamai2001a,Zhou2003a} and coordinated multipoint from the early 10s \cite{Gesbert2010a,Bjornson2010c,Boldi2011a}. Importantly, the spatial resolution of an array is not determined by the number of antennas but the array's aperture, so it is generally beneficial to spread out antennas, even if this also gives rise to spatial aliasing phenomena where signals coming from widely different directions cannot be separated \cite{massivemimobook}. {Non-uniform array geometries can provide better spatial resolution than uniform geometries \cite{Vaidyanathan2011a,Zhou2018a}.}

{We use the ELAA terminology to jointly describe a family of research topics that have previously been considered separately but all comply the following definition.

\begin{definition}
An ELAA consists of hundreds of distributed BS antennas that are jointly and coherently serving many distributed users.
\end{definition}

A list of different special cases is provided in Table~\ref{table:alternative-names}.
We believe that these special cases can, to a large extent, be jointly analyzed under the ELAA umbrella in the future.}

A consequence of using ELAAs is that the radiative near-field stretches many kilometers away from the array. Hence, the users will typically be in the near-field of the array, instead of the far-field as is traditionally the case, leading to non-stationary spatial channel properties \cite{Payami2012a,Amiri2018a}. In the far-field, the signal that reaches the array from a user is well approximated by a superposition of plane waves, each being determined by two parameters: a channel gain and an \ac{aoa}. In contrast, an array with an extremely large aperture can resolve not only the \ac{aoa} of a wave but also the distance it has traveled (e.g., the spatial depth) by exploiting the spherical shape of the wave and/or the channel gain differences to the antennas. It is also possible that some wave components are only  visible to a subset of the antennas in the array and blocked to the other antennas \cite{Liu2012c}. 
Hence, channel modeling is substantially harder when using ELAAs and involves many more parameters. { While conventional Massive MIMO benefits from channel hardening, where the small-scale fading average out since there are many antennas with similar channel gains, we cannot expect the same from ELAAs since well separated antennas have large gain differences \cite{Chen2017a}.
On the other hand, the great spatial resolution will likely make the channels to different users nearly orthogonal}
\cite{Payami2012a,Chen2017a,Hu2018a}, which is known as favorable propagation in the Massive MIMO literature \cite{massivemimobook}. Another consequence is that the transmission from the array cannot be illustrated as a beam, but rather as strong coherent signal amplification at distinct points in space, as illustrated in Fig.~\ref{fig:house}(b), while the antennas' signal components add non-coherently at most other places. {When aiming the beamforming at a particular point in space using well separated antennas, the signal components arrive from widely different directions but are phase-shifted to add constructively at the target point. Due to the different directivity, the}
 signal amplification gradually decays when leaving the target point, but irrespective of the antenna configuration it is always strong in a sphere around the target with radius $\lambda/8$ \cite{Bjornson2019a}. The volume that features signal amplification is typically of this size in the near-field while it can be much larger in the far-field, implying that closely located users can be spatially multiplexed with less mutual interference when using an ELAA.

\begin{table}
\begin{center}

\begin{tabular}{| l | l |} \hline
\textbf{Research direction} & \textbf{Special cases} \\
\hline
  Extremely large aperture array 
  & Cell-free Massive MIMO \cite{Ngo2017b}  \\ 
  &  Coordinated multipoint \cite{Boldi2011a} \\
  &  Very large aperture Massive MIMO \cite{martinez2014towards} \\
  &  Distributed MIMO \cite{Madhow2014a} \\
  &  Radio stripes \cite{Interdonato2018} \\
  & Network MIMO \cite{Venkatesan2007a} \\
  \hline
  Holographic Massive MIMO & Holographic RF system \cite{Prather2017a} \\
  & Holographic beamforming \cite{Black2017a} \\
  & Large intelligent surface \cite{Hu2017a} \\
  & Reconfigurable reflectarrays \cite{Hum2014a} \\
  & Intelligent walls \cite{Subrt2012a} \\
  & Software-controlled metasurfaces \cite{liaskos2018new} \\
  & Intelligent reflecting surface \cite{Wu2019a} \\
  \hline
\end{tabular}
\end{center}
\caption{The proposed research directions 1 and 2 collect many other research topics as special cases.} \label{table:alternative-names}
\end{table}

\subsection{Vision}

The grand vision of ELAAs is to provide orders-of-magnitude higher area throughput in wireless networks compared to what Massive MIMO with compact arrays can practically deliver.
The keys to reaching this goal are the even larger number of antennas and the distributed antenna deployment that reduces the average propagation loss and increases the spatial resolution (particularly, in the horizontal domain). There is a competing concept of ultra-dense networks (UDN) \cite{Hoydis2011c,Kamel2016a} that also relies on distributed antenna deployment, but each antenna box is then an autonomous BS that services its own exclusive set of users.
It is known that the throughput of UDNs is fundamentally limited by inter-cell interference \cite{Andrews2017a}. 
Cooperation between the distributed antennas can break this barrier as the number of antennas grows large \cite{ashikhmin2012pilot,BHS18A,Adhikary2017a,Neumann2018a,Sanguinetti2018a}, at least in theory. The ultimate goal of ELAAs is to deploy so {many coherently operating antennas} that all the users have mutually orthogonal channels, leading to a per-user throughput similar to that of an additive white Gaussian noise channel without any propagation loss \cite{Hu2018a,Bjornson2018b}.
In addition to enhancing mobile broadband services, which constitute the majority of the wireless traffic \cite{EricssonMobility}, the great spatial resolution of an ELAA can also be exploited for spatial multiplexing of an unprecedented number of machine-type communication devices. 

The mass production of smartphones has turned advanced antennas and transceiver equipment into a commodity. Similarly, the rapid development of integrated circuits (following Moore's law) has resulted in tiny processors that are incredibly computationally capable, essentially making the computational complexity of MIMO processing a non-issue \cite{Prabhu2017a,Perre2018a}. These facts can be exploited to achieve a cost-efficient deployment of ELAAs. The challenge lies in the interconnect of all the components in a system with thousands of cooperating antennas \cite{Perre2018a,Bjornson2019a}. One potential way to resolve this issue is to integrate the antennas and frontends into cables that can be attached to the facade of buildings and then connected to a baseband processor for coding and decoding of the data signals. This processor can either be physically close to the array or located in the basement or cloud.
This type of cable is called \emph{radio stripes} in \cite{Interdonato2018}. The concept is in many ways analog to the string lights that are used to light up (Christmas) trees; there is an electrical connector at one side of the cable, the cable can be bent and shaped as you like, no permission is required to deploy it, and the system continues to operate even if one component breaks down. The deployment in Fig.~\ref{fig:house}(b) can be achieved using multiple radio stripes, for example, one per floor in the building. There is no need for fine-tuning the physical location of the antennas when one has an abundance of them; in fact, an irregular deployment might even enhance the spatial separability of the users. {Another benefit of radio stripes is that a large number of antennas share a serial front-haul connection to the baseband processor, instead of having a separated connection per antenna, as in the pCell technology \cite{Perlman2015a} and initially considered for Cell-free Massive MIMO \cite{Ngo2017b}. This can greatly reduce the cost of the fronthaul infrastructure.} ELAAs will work efficiently even in a cellular deployment, with one ELAA of the type in Fig.~\ref{fig:house}(b) per cell, if MMSE processing methods are used to cancel inter-cell interference \cite{BHS18A,Bjornson2019c}.

Each antenna in an ELAA can be constructed using smartphone-grade hardware, instead of the industry-grade hardware used in contemporary BSs, and there are prospects of using even lower-grade hardware to cut cost \cite{Bjornson2015b,Zhang2018a}. It is like a divide-and-conquer approach to BS deployment; if we compare 1\,W transmission from a single antenna with $\frac{1}{M}$\,W-per-antenna transmission from $M$ antennas in an array, the latter can lead to $M$ times stronger received signal at the user due to the signal focusing, even if the total output power is 1\,W in both cases. Similarly, the uplink \ac{snr} is improved by collecting more signal energy with an ELAA. This is also where ELAAs differ fundamentally from UDNs, where each user is served only by the antenna that gives the largest channel gain. If we compare an ELAA and a UDN with the same antenna locations, the ELAA achieves stronger received signals and less interference since it coherently combines signals from many antennas to achieve the aforementioned divide-and-conquer gains. This has, for example, been demonstrated numerically for cell-free networks in \cite{Ngo2017b,Nayebi2017a,Bjornson2019c}. To achieve these gains, the antennas in the ELAA need to be phase-synchronized, which is more complicated to achieve in an ELAA than in a compact array. A synchronization algorithm based on over-the-air signaling is described in \cite{Interdonato2018}, indicating that this might be a solvable problem. There are also chip-scale atomic clocks that can be used to keep distributed radio-frequency (RF) components time synchronized \cite{CSAC45s}.

\subsection{Open Problems}

Since ELAAs provide a particular type of spatial channel correlation, the spectral and energy efficiency can be computed using known results on Massive MIMO with correlated fading \cite{Hoydis2013a,Adhikary2013a,BHS18A,massivemimobook}. The standard channel estimation, transmit precoding, and receive combining schemes can be readily applied---at least in theory. MR processing is very convenient for distributed deployments since there is no need to share channel knowledge between the antennas in the array \cite{Ngo2017b} but this benefit comes at a huge performance loss compared to using ZF or MMSE processing \cite{Nayebi2016a,BHS18A,Bjornson2019c}.
Hence, the first major open problem is to implement interference-rejecting precoding/combining schemes, such as ZF and MMSE processing, in a distributed or hierarchical way. When the channel gain variations are large over the array, it might only be worthwhile to let a part of the ELAA serve each user \cite{Amiri2018a,Buzzi2017a}. It is not the computational complexity that is the issue but the front-haul capacity requirements that would be extreme if thousands of antennas need to send their baseband samples to a common processing unit, possibly located at a remote location such as an (edge) cloud \cite{6882182,Perlman2015a}. It is important to develop theoretical distributed processing architectures, optimized resource allocation schemes, circuit implementations, and prototypes.

Channel modeling is another open problem for ELAAs. When using a compact planar array, as in Fig.~\ref{fig:AIR6468}(a), to serve \ac{los} users located in the far-field, the array response vector can be computed as a function of the azimuth and elevation angles of the incoming plane wave \cite[Section~7]{massivemimobook}. This is helpful when creating codebooks with beams that the users can select from, which is common for systems operating in FDD mode. It is challenging, if not impossible, to do the same for ELAAs. First, the users are in the near-field thus the wavefronts might not be plane. Second, the antennas can be arbitrarily deployed so the array response vectors are unknown, a priori, even for users located in the far-field. Hence, the modeling of channels, as well as the acquisition of the channel features in a particular deployment scenario, are important but challenging problems. Measurement campaigns and improved channel modeling based on physical properties and ray-tracing are needed. If the near-field behaviors turn out to be too complicated to model statistically, it will be necessary to define standardized evaluation scenarios with deterministic channels. Prototyping is also an important challenge and will naturally be more complicated and costly than in compact MIMO systems.

While an ELAA is very suitable for broadband and machine-type communications, it is hard to predict its impact on ultra-reliable low-latency communication services. {On the one hand, a large number of distributed antennas and an extreme spatial resolution can bring macro diversity against small-scale fading}, including the signal blockages that can occur at mmWave frequencies.
On the other hand, if we compare the conventional deployment in Fig.~\ref{fig:house}(a) with the ELAA in Fig.~\ref{fig:house}(b), the latter might be more susceptible to large-scale shadowing. Clearly, the reason for deploying BSs at elevated locations is to avoid that the signals are blocked by large objects. There is a risk that ELAAs will greatly improve the average user-performance, but degrade the worst-case performance, unless the deployment locations, antenna density, and operation are well optimized for ultra-reliability.

\begin{figure*}[t!]
	\centering
\begin{subfigure}[b]{\columnwidth}
	\begin{overpic}[width=\columnwidth,tics=10]{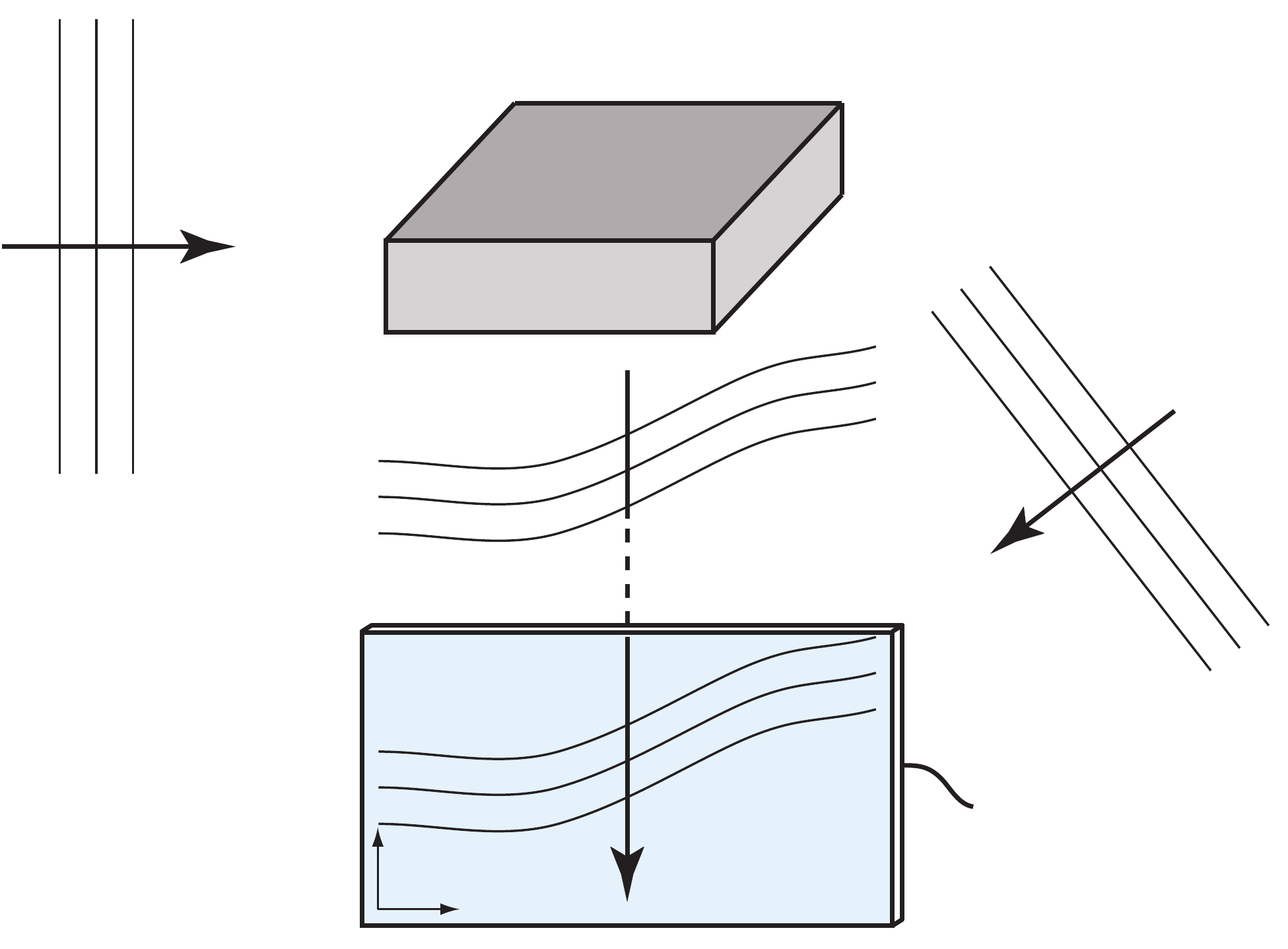}
  \put (0,75) {Illuminating wave}
  \put(39,68){Reflecting object}
  \put(83.5,47){Reference}
    \put(91,43){wave}
      \put(10,27.5){Reflected desired wave}
      \put(77,10){Recording}
      \put(77,6){medium}
      \put(34,17){$a(x,y)$}
       \put(34,4.5){$x$}
       \put(30.5,7){$y$}
\end{overpic}
        \caption{Holographic recording}
    \end{subfigure}%
    \begin{subfigure}[b]{\columnwidth}
	\begin{overpic}[width=\columnwidth,tics=10]{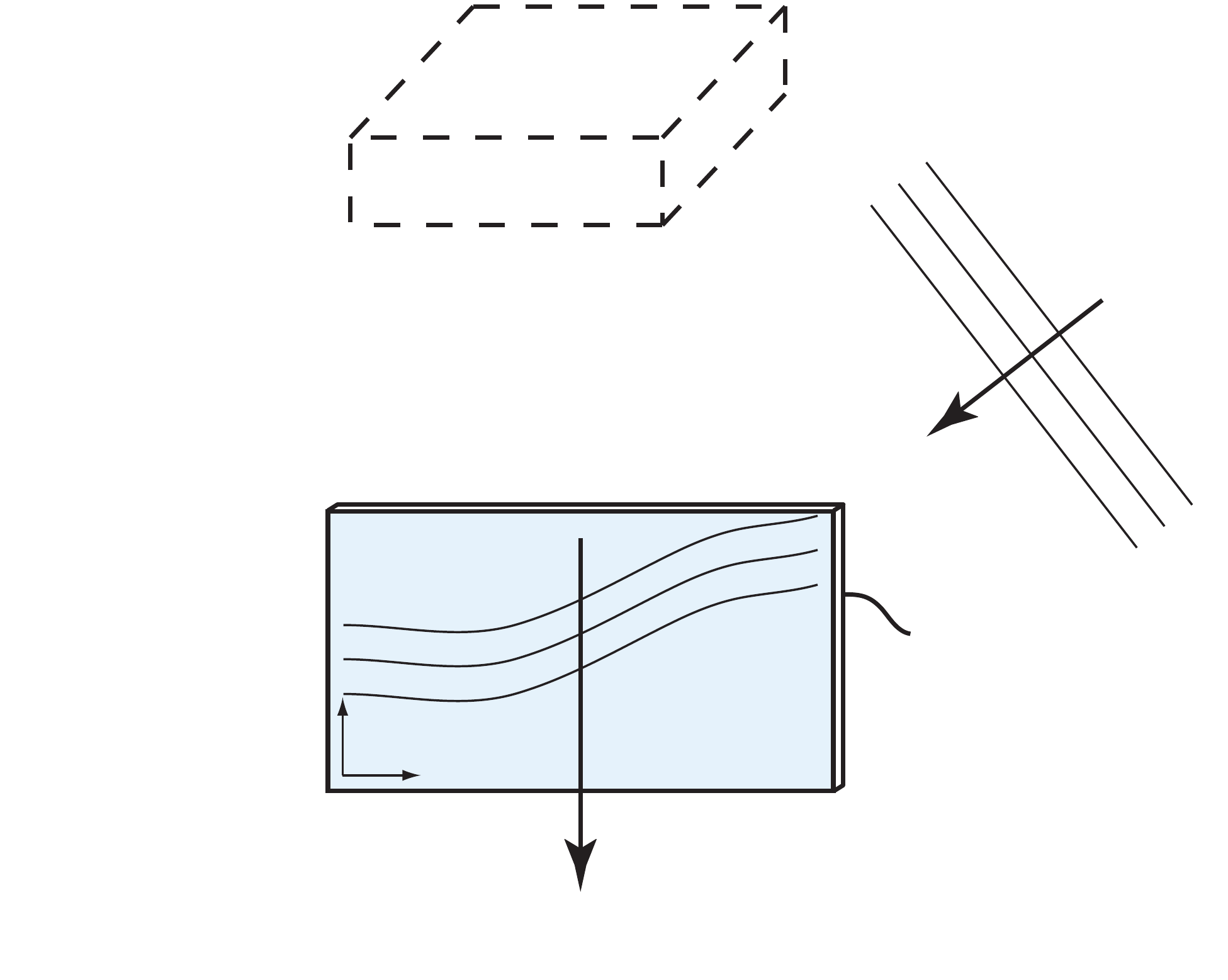}
  \put (30,57.5) {Virtual image}
  \put(85,61){Replica of}
    \put(87,57){reference}
    \put(93.5,53){wave}
    \put(20,3){Reconstructed desired wavefront}
    \put(33,32){$a(x,y)$}
      \put(75,27){Same medium}
      \put(75,23){acting as}
      \put(75,19){holographic}
      \put(75,15){display}
       \put(32.5,18){$x$}
       \put(29,20.5){$y$}
\end{overpic} 
        \caption{Holographic reconstruction}
    \end{subfigure} 
    \caption{Sketch of the two steps in optical holography. 
    (a) Holographic recording: the desired reflected wave $a(x,y)$ is mixed with a reference wave $e^{i (\alpha x + \beta y)}$ to record $|a(x,y) + e^{i (\alpha x + \beta y)}|^2$ when it reaches the recording medium.
 (b) Holographic reconstruction: the desired wave is reconstructed by illuminating the medium by a replica of the reference wave, turning the medium into a holographic display. Since the reconstructed wave is perceived identical to the original reflected wave, a virtual optical image of the reflecting object can be observed.}      \label{fig:holography} 
\end{figure*}
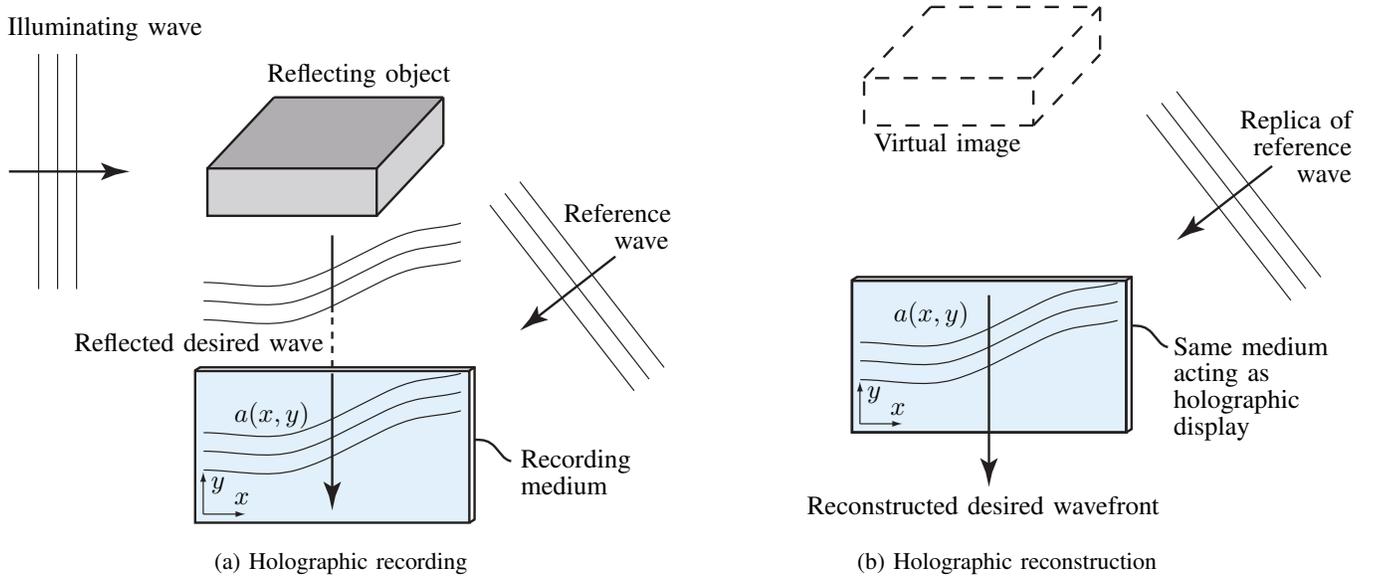

\vspace{5mm}

\section{Direction 2: Holographic Massive MIMO}

The capacity of Massive MIMO grows monotonically with the number of antennas \cite{massivemimobook} so it would be desirable to have nearly infinitely many antennas, but how can that be deployed in a practically viable way? The previous section described one option: physically very large arrays of classic antennas that are small and well separated to enable essentially invisible deployment. Another option is to integrate an uncountably infinite number of antennas into a limited surface area, in the form of a spatially continuous transmitting/receiving aperture.
This requires a radically new way of designing and analyzing antenna arrays; in fact, ``array'' might not be the right terminology anymore.
Research in this direction is taking place under the names of \emph{Holographic RF System}
\cite{Prather2017a}, \emph{Holographic Beamforming} \cite{Black2017a}, and \emph{Large Intelligent Surface} \cite{Hu2017a,Hu2018b}. When a spatially continuous aperture is being used to transmit and receive communication signals, we refer to it as \emph{Holographic Massive MIMO} since having an infinite number of antennas is the asymptotic limit of Massive MIMO.

How can we model communication when using a continuous aperture? Interestingly, the reception and transmission of an electromagnetic field over a continuous aperture is the defining characteristic of optical holography \cite{Goodman1968a}, where the received field is recorded and later reconstructed.
The recording medium (e.g., a photographic emulsion) only responds to the intensity $|a(x,y)|^2$ of the received field $a(x,y)$, where $x,y$ are the spatial coordinates on the detector, thus the phase is lost. The holographic recording/reconstruction process, illustrated in Fig.~\ref{fig:holography}, circumvents this issue by mixing the desired wavefront with a known reference wave. Suppose the plane wave $e^{i (\alpha x + \beta y)}$ is used as reference wave, then the combined wave $b(x,y)=a(x,y) + e^{i (\alpha x + \beta y)}$ reaches the medium, which will record
\begin{align} \notag
    |b(x,y)|^2 & = |a(x,y) + e^{i (\alpha x + \beta y)}|^2 \\
    &= |a(x,y)|^2 + 1 + 2 \Re \left( a(x,y)e^{-i (\alpha x + \beta y)} \right). \label{eq:holography:recording}
\end{align}
The last term in \eqref{eq:holography:recording} depends on the phase of $a(x,y)$ and the known phase of the reference wave, which means that the phase information has been implicitly recorded. To reconstruct $a(x,y)$, the transparent holographic display is illuminated by a replica of the reference wave, as shown in Fig.~\ref{fig:holography}(b), which yields the reconstructed electromagnetic field
\begin{align} \notag    e^{i(\alpha x + \beta y)} & |b(x,y)|^2 \\ \notag & = a(x,y) + a^*(x,y) e^{i2(\alpha x + \beta y)}  \\ &\quad+ \left(|a(x,y)|^2 +1  \right) e^{i(\alpha x + \beta y)}.  \label{eq:holography}
\end{align}
If the received wave is of finite angular extent, then a judicious choice of the reference wave separates the first component in \eqref{eq:holography} from the latter two. Hence, $a(x,y)$ is reconstructed and indistinguishable from the original field. In effect, optical holographic recording constitutes a distributed, homodyne receiver. If the display is not transparent but acts as a mirror in the reconstruction phase, we can instead illuminate it by the conjugate $e^{-i (\alpha x + \beta y)}$ of the reference wave, which leads to emitting the conjugate wavefront $a^*(x,y)$ in the opposite direction.
Analogously, in wireless communications, the illuminating wave is a pilot waveform emitted by a user device and the reflecting object is the propagation environment, and the reference wave might be generated inside the surface. By emitting the conjugate wavefront from the surface, we can effectively transmit back to the user device using MR precoding (also known as conjugate beamforming).

\begin{figure*}[t!]
	\centering
\begin{subfigure}[t]{\columnwidth}
	\begin{overpic}[width=\columnwidth,tics=10]{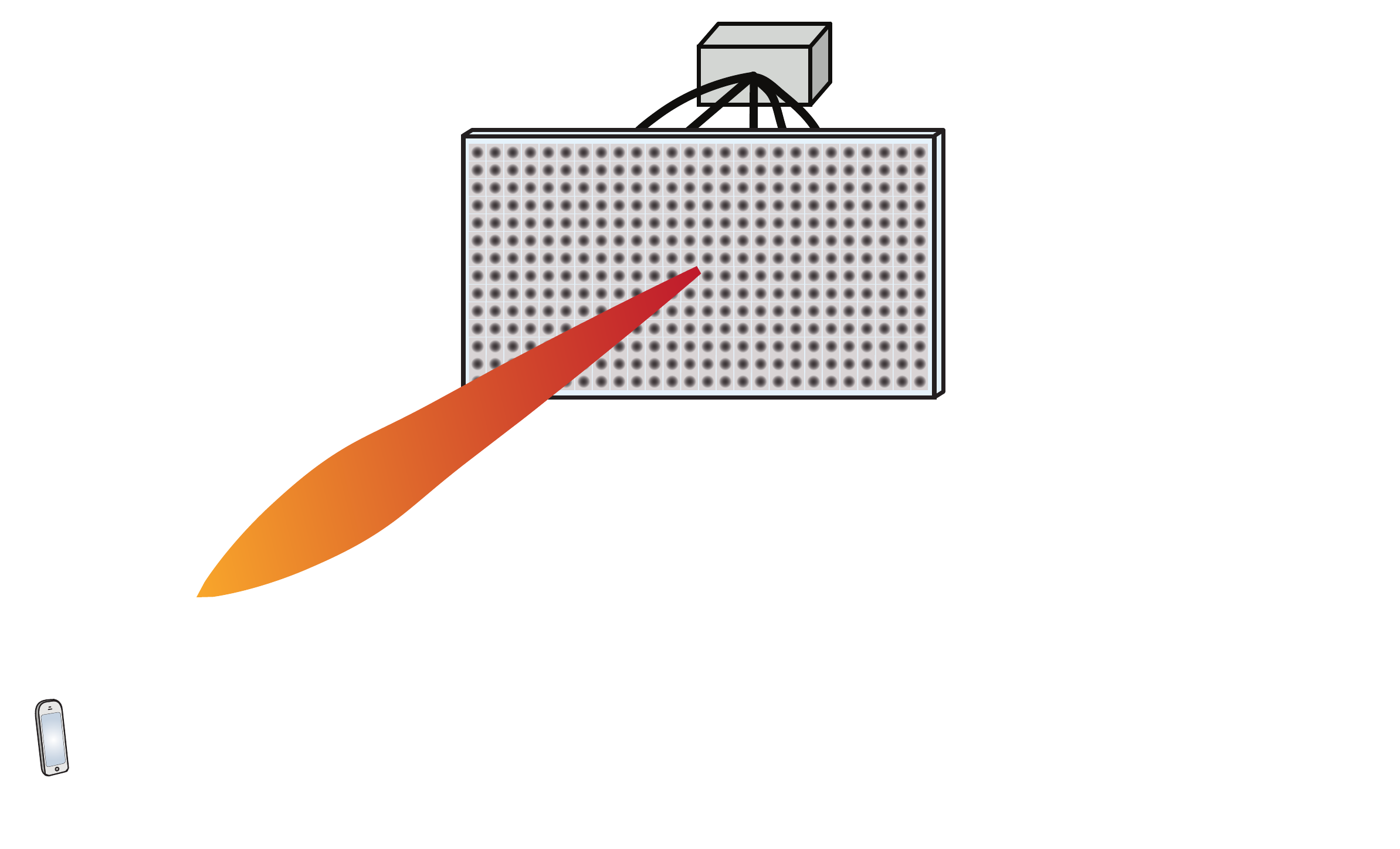}
      \put(62,55){RF signal generator}
      \put(7,6){Receiving user}
      \put(45,28){Surface with}
      \put(45,24){radiating elements}
\end{overpic}
        \caption{Contiguous surface with an RF signal generator at the backside.}
    \end{subfigure}%
    \begin{subfigure}[t]{\columnwidth}
	\begin{overpic}[width=\columnwidth,tics=10]{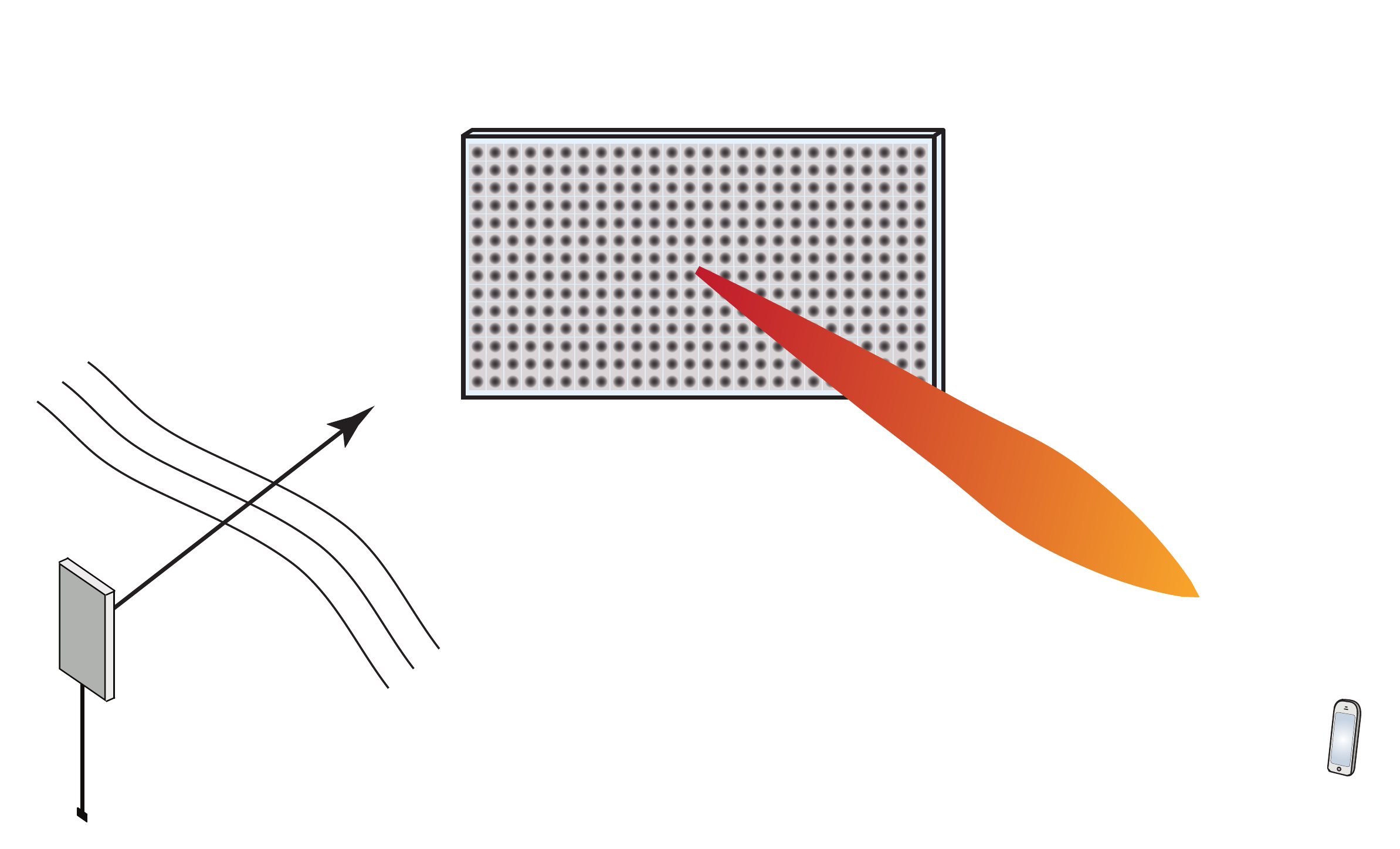}
      \put(68,6){Receiving user}
      \put(10,8){RF transmitter}
      \put(33,28){Surface with}
      \put(33,24){reflecting elements}
\end{overpic} 
        \caption{Contiguous surface with an RF signal generated at another location.}
    \end{subfigure} 
    \caption{Two examples of Holographic Massive MIMO. (a) The RF signal is generated at the backside of the surface and propagates through a steerable distribution network to radiating elements that generate a beam. (b) The RF signal is sent from another location and the metasurface reflects it using steerable elements that generate a beam.}      \label{fig:surface} 
\end{figure*}
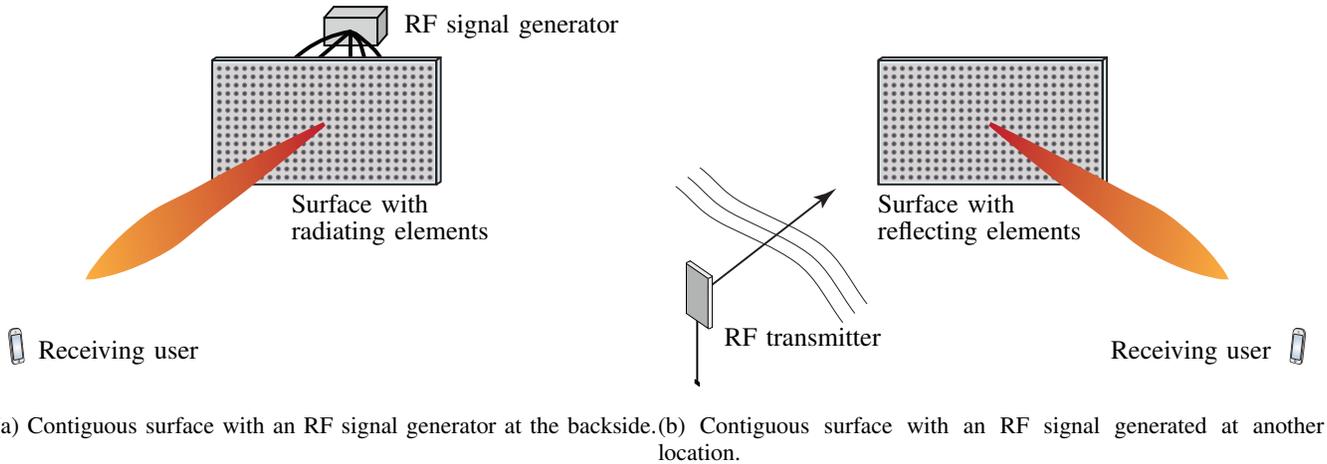

Two approaches are currently being taken to approximately realize a continuous microwave aperture:
\begin{enumerate}
\item The first approach uses a tightly coupled array of discrete, active antennas \cite{Prather2017a}. {This can be implemented using a dense array of conventional fraction-of-$\lambda$-spaced antennas, connected to RF chains, but the result will be costly and bulky.} Alternatively, the RF signals are generated by mixing two optical signals whose frequency difference equals the desired RF carrier frequency \cite{Bai2014a}.

\item The second approach mixes an RF reference signal with a large number of nearly passive reflecting elements having electronically steerable reflection parameters \cite{Huang2005a,Hum2014a}, {known as a reconfigurable reflectarray}. The specific implementation described in \cite{Black2017a} uses a single RF port on the backside of the surface, which is connected to an electronically steerable RF distribution network with radiating elements that emit the wavefront; {see Fig.~\ref{fig:surface}(a).} Using the holographic terminology, the RF port generates the reference wave and the distribution network is the holographic display. If multiple RF ports are connected to the same distribution network, a superposition of ``beams'' can be transmitted and received, to enable spatial multiplexing. This is conceptually similar to the hybrid architectures considered in the mmWave literature (see \cite{Xiao2017a} and references therein), but differs in the use of holographic recording for beamforming design instead of the complicated beam-training procedures commonly considered for hybrid beamforming.
\end{enumerate}

{Another embodiment of reconfigurable reflectarrays is \emph{software-controlled metasurfaces}, which are contiguous surfaces that are not actively emitting RF signals, but consist of many discrete ``meta-atoms'' with electronically steerable reflection properties (e.g., shift in phase, polarization, and amplitude) \cite{Subrt2012a,liaskos2018new}. Hence, when illuminated by an RF signal sent from another location, the metasurface can control the reflections and, for example, beamform the RF signal in a preferable way; see Fig.~\ref{fig:surface}(b). In principle, the metasurface can be reconfigured to behave as an arbitrarily shaped mirror in the given frequency band. 
Metasurfaces are also known as
\emph{intelligent reflecting surfaces} \cite{Wu2018a,Wu2019a} and there are several variations on these names  \cite{Huang2018a,Yu2019a,Renzo2019a,He2019a}. Note that the key difference from the second approach, described in the previous paragraph, is that the RF signal is not generated inside or close to the surface, but relatively far away. We can nevertheless provide the following joint definition.}

\begin{definition} 
In Holographic Massive MIMO, an antenna surface with (approximately) continuous aperture is used for communication. The surface either actively generates beamformed RF signals or control its reflections of RF signals generated at other locations.
\end{definition}

Our focus here is the motivation for, and the potential benefits of deploying a continuous aperture, and the associated research problems.

\subsection{Vision}

There are three main reasons to consider Holographic Massive MIMO. First, a continuous aperture system may become a more attractive way to achieve high spatial resolution in wireless communications than having a very large number of conventional discrete antennas, each with dedicated hardware (this is definitely true at optical wavelengths). The number of RF ports in Holographic Massive MIMO can be equal to the number of signals to spatially multiplex. Second, the continuous aperture enables the creation and detection of electromagnetic waves with arbitrary spatial frequency components, without undesired side-lobes. In other words, the same surface can be reconfigured to work flawlessly on any carrier frequency since the classic spatial aliasing effects \cite{massivemimobook} that occur when having sub-critically spaced antennas are alleviated by the continuous aperture. Third, a continuous-space system should be more tractable to analyze than a large-scale discrete-space system; generally, integrals are more convenient than discrete summations.

The vision is that Holographic Massive MIMO can be integrated into any surface, including walls, windows, and even fabrics. We can eventually be surrounded by surfaces that can emit and receive any electromagnetic waves, while remaining  aesthetically appealing. The extreme spatial resolution enables incredibly low transmit powers and unprecedented spatial multiplexing capabilities \cite{Hu2018a}. As a theoretical tool, Holographic Massive MIMO can be also used to investigate the true limit behavior of Massive MIMO communication systems.\footnote{The existing asymptotic analysis for Massive MIMO with uncorrelated \cite{Marzetta2010a} and correlated fading  \cite{BHS18A} use simplified channel models where the receiver will asymptotically capture more energy than was transmitted \cite{Bjornson2018b}, which is clearly impossible. Although these results manifest the performance scaling for a practical number of antennas, they cannot be used to study the asymptotic performance limit. Hence, further work is required and \cite{Hu2018a} provides one step in this direction.} 
It is clearly simpler to analyze a system comprising a continuum of antennas instead of a countably infinite number of randomly located antennas.

The potential use cases for Holographic Massive MIMO go far beyond conventional communications. A classic planar antenna array, utilizing $\lambda/2$-spacing on a Cartesian grid, creates on transmission and detects on reception a discrete set of ordinary (propagating) plane waves, whose three wavenumber components each have a magnitude less than $2\pi / \lambda$. Keeping the same physical array dimensions, but simultaneously adding more antennas and reducing the spacing brings additional, evanescent, plane-waves into play, that, parallel to the array have wave-number magnitude greater than $2\pi / \lambda$, and perpendicular to the array, have an imaginary-valued wave-number. Perpendicular to the array, evanescent waves decay exponentially fast in space, and they carry only reactive (imaginary) power. For this reason, evanescent waves do not participate in conventional far-field wireless communications, and replacing a conventional discrete array with a holographic array of the same physical size would add no additional degrees of freedom. Notwithstanding, evanescent waves are critical to some unconventional communication schemes; for example, extreme near-field communication, wireless power transfer via resonant evanescent wave coupling, and super-directive antenna arrays. These examples are discussed in further detail below.

In the extreme near-field, evanescent waves contribute significantly to the electromagnetic field. An example of extreme near-field communication is that of an implantable neural transmitter \cite{Yin2013b}, which continuously digitizes and transmits 100 neural channels (48\,Mb/s) of data through the scalp to an external receiver a few meters away, using 3.2/3.8\,GHz FSK modulation. It is interesting to speculate on the possibility of increasing the number of neural channels by a factor of ten, one hundred, or one thousand---present-day communication theory seems inadequate. Since one wavelength is equivalent to the radius of the brain, any MIMO approach would, of necessity, entail closely spaced antennas.

Resonant evanescent wave coupling has been used for wireless power transfer between a pair of coils, at the  10\,MHz carrier frequency (30\,m wavelength): 60\,W of power transfer was achieved over a range of two meters with 40\,\% efficiency \cite{Karalis2008a}. The near-field is then dominated by evanescent waves. The actions of the receiving coil, tuned to resonance, and having a high Q, force the evanescent waves, contrary to their nature, to transport real power, with high efficiency.

Super-directivity is another phenomenon that is seldom considered in communications. By placing antennas close together to realize super-directivity, in principle, one can achieve array gains far in excess of what conventional arrays and MR processing achieve \cite{Foschini1995a,Foschini1998b}: ``For example, consider a transmitting horn antenna, with an aperture about 10 wavelengths on a side, located in outer space roughly aimed at the earth. With a one wavelength diameter super-gain antenna on the earth it is possible to receive virtually all of the power radiated by the horn antenna.''

\subsection{Open Problems}

A general challenge presented by Holographic Massive MIMO is that the progress depends on the availability of researchers who are conversant in both communication theory and electromagnetic theory \cite{Franceschetti2017a,Marzetta2019a,Hu2018a}. Holographic Massive MIMO is investigated in \cite{Hu2018a,Hu2018b} for communication and positioning, respectively, under \ac{los} propagation. An extension to scattering propagation would require that the conventional i.i.d.~Rayleigh model for small-scale fading is discarded and replaced by a spatial correlation model, but none of the conventional statistical models for uniform linear arrays with discrete antennas are directly applicable.\footnote{Physics dictates that the small-scale fading satisfies the homogeneous wave equation. If the random field is also required to be spatially stationary, then the power density spectrum for the small-scale fading exists on the surface of an impulsive sphere of radius $2\pi / \lambda$. Under this model, there is a plane-wave spectral representation for the random field, which yields an efficient method of computing the joint likelihood of the continuum of noisy measurements \cite{Marzetta2018a}.}
Further research on channel modeling, as well as channel estimation that utilizes its inherent spatial correlation structure, is highly needed.
The accurate calculation of the communication performance, in general, requires elaborate numerical electromagnetic computations, but these details should be of little concern to the communication theorist: the mere existence of the plane-wave representation, and the impedance kernel, provides the essential physics for the formulation of physics-based communication theory. {Nevertheless, a general framework for exact or approximate performance analysis remains to be created, and there is a risk that simplified models lead to incorrect results.}

{The practical implementation of Holographic Massive MIMO has started \cite{Hum2014a,Black2017a} but more remains to be done. For arrays that generate the RF signals internally, it is important to have multiple RF ports connected to the same surface, so that users can be spatially multiplexed.} So far, the transmission from a continuous aperture has been approximated by discrete patches and this leads to mutual coupling \cite{Hu2018c}, which has to be accounted for in the computation of radiated power. Physically, the action of feeding current into a patch performs work against the field created by the patch, itself, and the field created by the rest of the patches. As a consequence, the power is not equal to a simple integral of magnitude-current-squared but rather is equal to a quadratic form of current with respect to a positive-definite impedance kernel. {Hence, there are non-trivial factors to consider in the signal generation and beamforming design.} Many important open problems in the intersection between hardware properties and electromagnetic waves can be found in this area. 

{For surfaces that reflect RF signals generated elsewhere, many implementation concepts with discrete reflection elements have been developed \cite{Huang2005a,Yang2016a,liaskos2018new}. Several potential use cases have been identified \cite{liaskos2018new,Wu2019a,Renzo2019a} but it remains to find a killer application---something that cannot be done with existing technology. For example, it has been shown that metasurfaces can provide range extension \cite{Wu2018a,Huang2018a}, but the gains over conventional relaying are not convincing \cite{Bjornson2019d}. Efficient protocols and algorithms for channel estimation and fast reconfiguration of the reflection properties need to be developed; the narrower the reflected beams are, the more important the estimation and configuration accuracy will be. Since the phase-shifts are frequency-flat, another challenge is to operate the surfaces over frequency-selective channels and to also transmit efficiently to users in the near field.}

{To summarize, Holographic Massive MIMO is an exciting and challenging field of research, where electronically active surfaces are designed and utilized to receive, transmit, and reflect arbitrary waveforms. The continuous aperture can be employed in conventional communication applications to form sharp beams without side-lobes and potentially operate close to the asymptotic limits of Massive MIMO, in terms of both channel capacity and energy efficiency. The surfaces can also enable unconventional applications of wireless, such as highly efficient wireless power transfer.}

\vspace{5mm}

\section{Direction 3: Six-Dimensional Positioning}
\label{sec:positioning}

An important application of any cellular communication system is the ability to spatially locate its users \cite{Gustafsson2005}. While the positioning requirements were traditionally determined by mandates to localize emergency calls, every new generation of cellular networks has provided new positioning opportunities \cite{Peral-Rosado2018}: 2G was limited to cell-ID and coarse timing measurements, with horizontal accuracies around 50-200\,m. In 3G, the larger bandwidths made time-difference-of-arrival measurements more accurate, while in 4G, the positioning performance was boosted through dedicated reference signals. Techniques such as carrier aggregation, MIMO, and device-to-device communication all improved the positioning performance in successive LTE releases, with accuracies now being in the 10\,m range for 10\,MHz carriers \cite{del2018position}. Generally, {positioning is an application of statistical signal processing where a position is computed} through the fusion of wireless communication signals with external information, for example, obtained by the \ac{gnss} or barometers \cite{mensing2010hybrid}.

To break the 10\,m accuracy barrier and support new applications, disruptive technologies are needed \cite{laoudias2018survey}. Let us  first recap the fundamentals of positioning. A user's position is generally obtained through a two-stage process \cite{zekavat2011handbook}:  measurements are first collected and then a position estimate is computed. The accuracy is thus fundamentally limited by the quality of the underlying measurements. {The bandwidth plays an important role since it determines the sampling rate and the ability to resolve multipath components in the frequency domain.}
For time-based measurements, commonly used from 2G until 4G, Fisher information analysis reveals that the measurement variance  is inversely proportional to the \ac{snr} and the square of the signal bandwidth, while the resolution (the difference in arrival time between two paths that can still be distinguished) is limited to the inverse of the bandwidth \cite{shen2010fundamental}. Hence, high resolution and high accuracy can only be obtained when a large bandwidth is available. The accuracy is also improved with more BSs, with three-dimensional positioning based on distances requiring at least four \ac{los}  measurements (due to lack of synchronization between the user and the BSs). 

MIMO has thus far only played a marginal role in wireless positioning since small arrays provide little benefit. With the realization of Massive MIMO, new dimensions and opportunities for positioning open up. In 5G, we might simultaneously have more antennas and more bandwidth than in previous technology generations. This combination will be instrumental when superior positioning precision is needed.

\subsection{Vision}

The ultimate goal of positioning is to precisely estimate not only the three-dimensional spatial location, but also the three-dimensional orientation of the user device, represented by roll, pitch, and yaw. We refer to this as \emph{Six-dimensional positioning}. Our vision is that this can be achieved using antenna arrays with many antennas, particularly when combined with high carrier frequencies.

First of all, the use of arrays at the BSs enables estimation of new physical quantities, namely the \ac{aoa} (in uplink) and \ac{aod} (in downlink) \cite{Garcia2017}. A Fisher information analysis shows that the variance of the \ac{aoa} estimates decreases cubically with the number of antennas, while the variance of the \ac{aod} estimates decreases quadratically with the number of antennas \cite{Garcia18}.
The main reason is that the beamwidth of the array decreases with the number of antennas \cite{massivemimobook}, which dictates how small angular differences can be resolved.\footnote{To be more precise, the beamwidth of the main-lobe is determined by the length of the array relative to the wavelength. Hence, adding more antennas will decrease the beamwidth (i.e., increase the spatial resolution) if the antenna spacing is constant so that the array size is growing. If we are instead adding more antennas into a fixed array size, it is instead the shape of the side-lobes that changes; see \cite[Sec.~7.4]{massivemimobook} for details.} Here, the number of antennas should be interpreted along the direction we are interested in; that is, in the horizontal direction for azimuth and the vertical direction for elevation.\footnote{The beamwidth of the main lobe can be very different in the azimuth and elevation domains, if a one-dimensional or strongly rectangular two-dimensional  array is used.}
For example, the 64-antenna array depicted in Fig.~\ref{fig:AIR6468}(a) has eight dual-polarized antennas in the horizontal direction and four in the vertical direction, none of which is particularly massive. 
Arrays with many more antennas in the horizontal/vertical directions (i.e., larger apertures) are desirable to further improve the positioning accuracy in the future. This can be implemented on lower frequencies using the ELAA concept or by increasing the carrier frequency to squeeze more antennas into the same physical size. The use of fully digital arrays  enables advanced signal processing for positioning.
The potential for high-quality measurements obtained by antenna arrays and new signal processing methods can, in turn, reduce the BS density that is required to satisfy a given positioning accuracy \cite{guerra2018single}. In principle, a user can be positioned in two dimensions using a single \ac{aoa} or \ac{aod} measurement with a two-dimensional array, since a single direction intersected with a plane yields a unique point. Even when operating at lower frequencies, the spatial richness of Massive MIMO signals allows for precise positioning using data-driven methods \cite{vieira2017deep,Decurninge2018}.

\begin{figure}
	\centering
	\begin{overpic}[width=\columnwidth,tics=10]{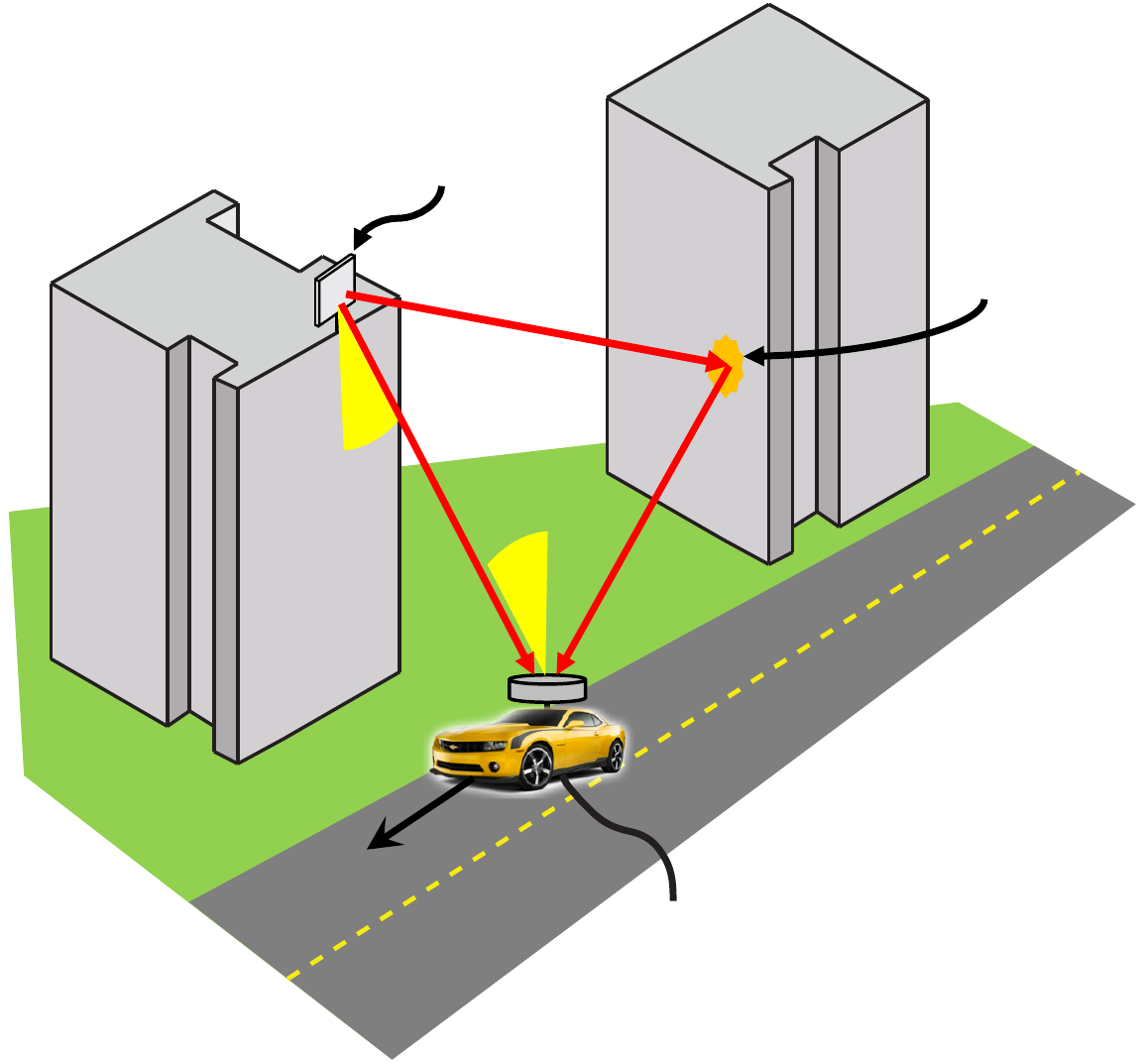}
  \put (15,78) {Massive MIMO array}
  \put(37.3,67){Reflected path}
    \put(84,73){Incidence}
    \put(84,69){point}
      \put(37,56){LoS path}
      \put(45,39){\footnotesize \rotatebox{90}{AoA}}
      \put(30.5,55){\footnotesize \rotatebox{90}{AoD}}
    \put(53,11){User with unknown position,}
    \put(53,7){orientation, and clock bias}
\end{overpic} 
	\caption{Precise estimation of a user's position and orientation is possible by exploiting the spatial resolution provided by large MIMO arrays. Under favorable conditions, multi-path can be exploited to perform mapping and synchronization.}
	\label{fig:loc1}
\end{figure}
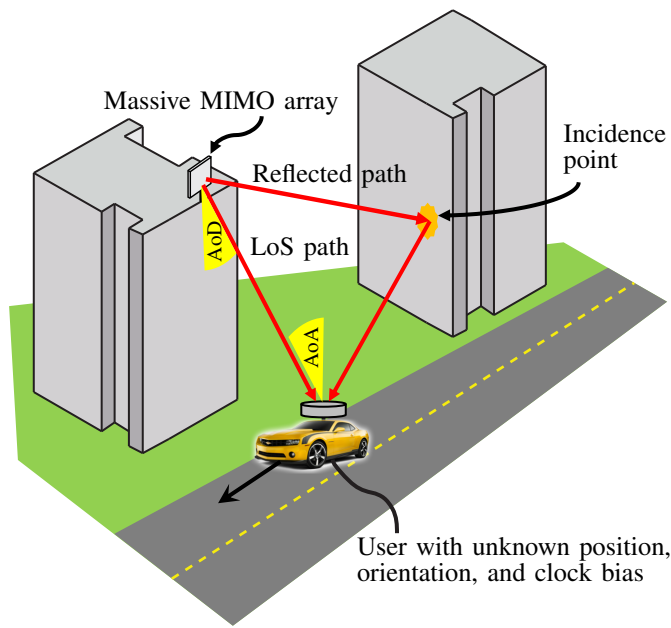

Second, if the user is also equipped with an antenna array, then both \ac{aoa} and \ac{aod} can be estimated simultaneously  in either uplink or downlink \cite{Shahmansoori2018}, as shown in Fig.~\ref{fig:loc1}. This allows the orientation of a user (i.e, roll, pitch, and yaw) to be also estimated, leading to six-dimensional positioning. Estimation of the orientation parameters in addition to an absolute location has benefits for several applications, such as positioning of \ac{uavs}, ground robots, in factory automation, or for automated vehicles \cite{wymeersch20175g}, 
as well as new disruptive applications in the contexts of virtual and augmented reality, gaming, social networking, etc. \cite{marchand2016pose}. 

Third, the fusing angle and delay information can resolve paths in the delay or angle domains, providing a richer picture of the propagation environment \cite{Witrisal2016}. In particular, at mmWave frequencies, where channels are relatively sparse and closely related to the geometry of the environment, it becomes possible to use the communication signal as a type of radar sensor, which provides sensing of nearby objects in terms of angles and distances \cite{guidi2016personal}. Just as modern automotive radars can produce three-dimensional maps of the environment, so can a Massive MIMO system. These maps can then be harnessed by other users to improve their position or orientation estimates \cite{laoudias2018survey}. Hence, high resolvability in delay or angle turns multi-path from  a foe to a friend, in contrast to conventional thinking \cite{gentner2016multipath}. This is further discussed in the next section.

Finally, the propagation paths provide geometric constraints with more observables than unknowns, which leads to opportunities to synchronize the user's local clock/oscillator to the BS without dedicated synchronization signals. Such joint time synchronization and positioning have been considered both for uplink MIMO \cite{koivisto2017joint} and for downlink MIMO \cite{wymgarkim:18}, even when the \ac{los} path is blocked.

\subsection{Open Problems}

To realize the vision of Massive MIMO positioning, several problems remain to be solved. The most urgent issues are the availability of tractable but sufficiently rich geometric position-dependent channel models at different frequencies and a set of common use cases for evaluation purposes. In contrast to models designed for communication applications, the positioning community requires ground truth of the BS, user, and environment, with an accuracy far exceeding that of the desired positioning accuracy. Wireless channels that are good for communication may be challenging for positioning and vice versa. The impact of different material properties should be studied for positioning, especially as the spread in the angle and delay domains will affect the performance. Ideally, a reference database  (e.g., as in \cite{torres2016providing}) of Massive MIMO signals at lower (below 6\,GHz) and higher (above\,28 GHz) frequencies should be made available to the community, in order to test positioning algorithms in a fair and standard way, to guarantee real and quantifiable progress in the field. 

Second, the interaction between positioning and communication should be considered \cite{koivisto2017high}. While most communication systems are using OFDM, the preferable distribution of pilot carriers and pilot beams will be different for positioning applications. Design of waveforms suitable for positioning of many users, either in uplink or downlink, using either a single BS or multiple BSs requires further study. When positioning is carried out by user devices, the computational complexity is also an important consideration. Algorithms that can perform positioning and {create maps (so-called mapping)} in real-time are challenging \cite{leitinger2018scalable}, due to the nonlinear measurements in Massive MIMO. \ac{slam} algorithms for multi-user problems is still an active area of research \cite{battistelli2013consensus}. 

Third, even when channel models and algorithms are available, many engineering issues remain. These include the need for synchronization among BSs and possibly with users. Moreover, the antenna positions and their orientation need to be calibrated to an extent demanded by the positioning application. This stands in contrast to the reciprocity-based beamforming advocated in the Massive MIMO literature \cite{Marzetta2016a,massivemimobook}, where no prior information of the antenna positions/orientations is needed \cite{Bjornson2016b}; the BS simply estimates the channel from uplink pilots and uses it for beamforming without extracting any geometric features. The impact of antenna position/orientation misalignment on the positioning accuracy of large arrays is an open problem. If a fully digital array is used, the misalignment can potentially be estimated and compensated for by signal processing.
It is worthwhile to note that generally, the positioning requirements are more stringent than the communication requirements (e.g., a 10\,ns error is tolerable for many communication applications, but leads to positioning errors of 3\,m or more). 

Finally, with the ability to position users and map the environment come concerns related to privacy and security, and even the dual use or misuse of positioning technologies \cite{lohan20185g}. While such concerns are present in any positioning application, the  prevalence of precise Massive MIMO positioning and its use in safety-critical applications such as \ac{uavs} demands special care from the community \cite{altawy2017security}. Where information will be stored and who will have access to it, as well as what operators and application developers will have access to, is the subject of both research and regulation.

\vspace{5mm}

\section{Direction 4: Large-scale MIMO Radar}

\begin{figure}[t!]
	\centering \vspace{-2mm}
	\begin{overpic}[width=\columnwidth,tics=10]{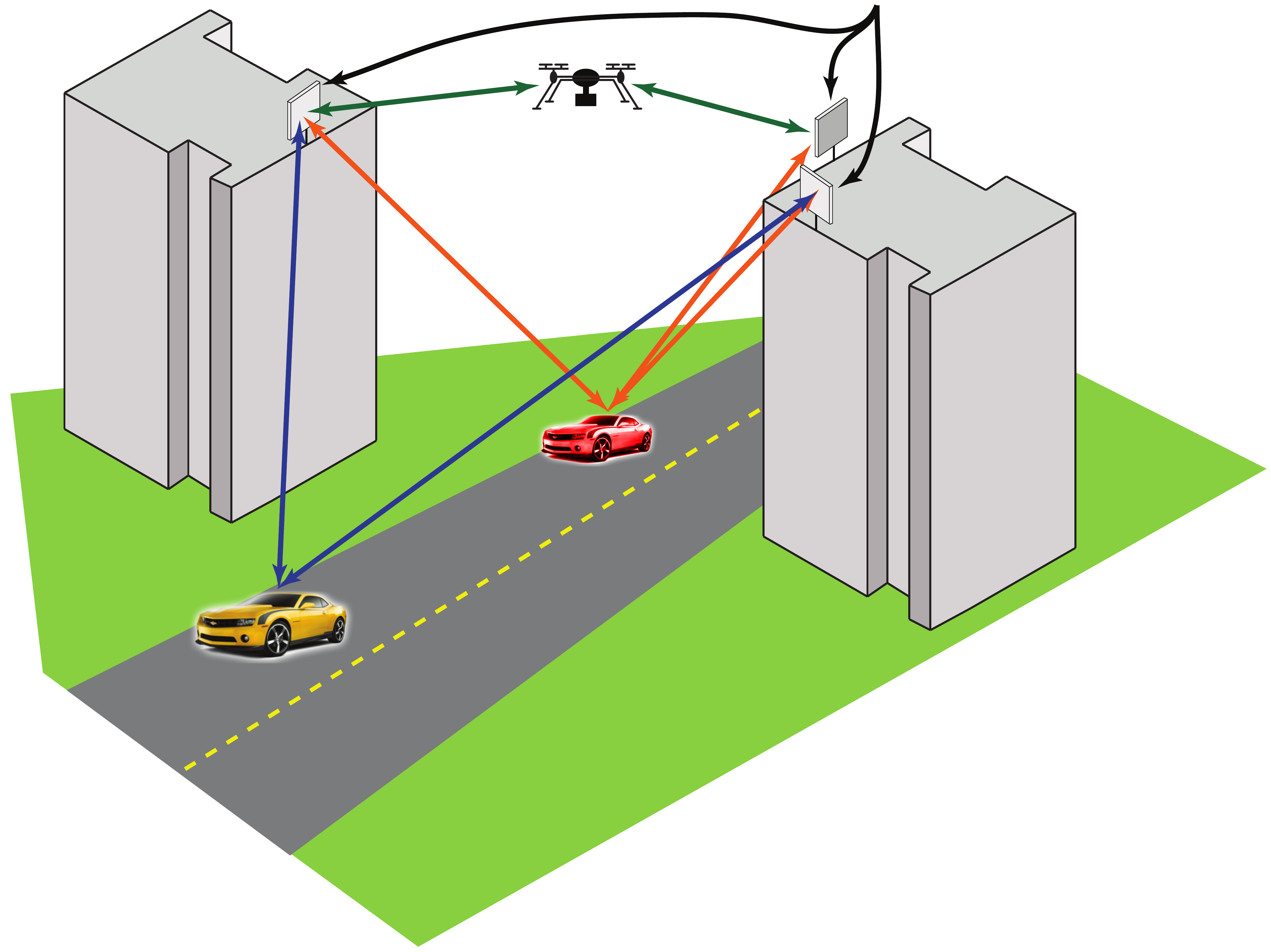}
	  \put (40,76) {Large-scale MIMO radars}
\end{overpic} 
	\caption{Illustration of a radar system consisting of three Large-scale MIMO radars that are used to detect and track three moving targets. The arrows indicate different probing signals and reflections.}\label{fig:MIMOradar}
\end{figure}

Radar stands for \emph{RAdio Detection And Ranging} and is a system that uses electromagnetic probing signals to detect and measure parameters (such as location and speed) of reflecting objects, called targets. The radar station transmits electromagnetic waves into a search area and receives the echo signal reflected from a target. By applying signal processing algorithms to the received signal, the target can be detected, and its parameters estimated. The conventional configuration is called \emph{monostatic} and refers to a radar in which the transmitter and receiver are co-located. The term monostatic is used to distinguish it from a \emph{bistatic} radar (wherein the transmitter and receiver are separated by a distance comparable to the expected target distance) and \emph{multistatic} radar (consisting of multiple monostatic or bistatic radars). An example of a multistatic radar is given in Fig.~\ref{fig:MIMOradar}, where there are three radar components and three targets.

While radar is traditionally used for military applications, aerospace security, and weather services, new applications have recently appeared in the automotive sector \cite{Patole2017}, medical devices \cite{Nosrati2019}, and civil engineering \cite{Pajewski2013}, to name only a few.
Analog phased arrays have been used in military radar systems, both at the transmitter and receiver ends, since the early 70s thanks to their capabilities of steering a beam in any direction across the search area (see Fig.~\ref{fig:beamforming}(a)). In addition to angular beamforming, they have been also used to improve the estimation accuracy of target parameters, using high-resolution methods such as MUSIC and maximum likelihood \cite{Stoica98}.

There is a strong connection between fading channels in communications and the fluctuations in the received radar signals.
Both experimental and theoretical results have demonstrated that a radar target provides a rich scattering environment that yields tens of dB or more of radar cross section (RCS) fluctuations\footnote{The RCS function of the target basically represents the amount of energy reflected from the target as a function of its aspect. The term ``aspect'' refers to the angle at which the target is inclined toward the incoming radar wave.} and this makes it behave as a fading channel \cite{Fishler2004}. In communications, the channel fluctuations can be exploited to achieve a \emph{diversity gain}; that is, to combat the fading by observing independent fading realizations, which are unlikely to be all simultaneously small.\footnote{The channel hardening property of Massive MIMO is the massive form of spatial diversity.} Inspired by this concept, the MIMO radar in \cite{Fishler2004} consists of a fully digital transmit array with widely spaced elements (distributed over a large geographical area such that each sees a different aspect of the target) and capitalizes on the independent RCS fluctuations to obtain a more uniform distribution of the received signal. This can be exploited to obtain better target detection performance \cite{Fishler2004a} and improved parameter estimation \cite{Haimovich2008}, as compared to phased arrays that perform analog beamforming. 

In addition to diversity, another important feature of MIMO in communications is the spatial multiplexing of data streams, each having different spatial directivity. This is the most efficient way to improve the capacity in multi-user MIMO and, particularly, Massive MIMO \cite{massivemimobook}.
In radar applications, the same capability can be exploited to spatially multiplex correlated or uncorrelated probing signals and thus improve system performance (compared to phased arrays) in several ways \cite{Stoica2007}: 1) increasing the number of targets that can be uniquely identified (up to $M$, where $M$ is the number of transmit antennas) \cite{Li2007}; 2) achieving high spatial resolution and improving interference rejection capabilities \cite{Bliss2003}; and 3) generating different beampatterns with superior performance \cite{Stoica2007a}.

For all these reasons, the MIMO radar paradigm has received great attention from researchers and has rapidly evolved from its original concept in \cite{Fishler2004}. The term MIMO radar is nowadays referring to any fully digital radar system employing multiple transmit waveforms (usually referred to as \emph{waveform diversity}) and having the ability to jointly process signals received at multiple receive antennas. Based on the array configurations used, MIMO radars can be classified into two main types. The first type uses widely separated antennas (so-called \emph{distributed} MIMO) to capture the spatial diversity of the target's RCS \cite{Haimovich2008}. The second type employs arrays of closely spaced antennas (so-called \emph{co-located} MIMO) to coherently combine the probing signals in certain points of the search area \cite{Stoica2007}. Hybrid configurations are also possible \cite{li2007a,Xu2007,Hassanien2010}.

Some of the theoretical findings for MIMO radars have already been validated in experiments, which have been conducted over the years by different research groups, see \cite{Robey2004,Frazer2009,Bliss2009,Frazer2017} and references therein. 
Despite this, MIMO radar is still a controversial topic. The hype surrounding it in academia, since the early papers, has come with a fair amount of skepticism from radar engineers. This has slowed down its adoption in practice, where analog phased arrays are still being used in most applications. The benefits and pitfalls of MIMO radar are reviewed in \cite{Daum2009}, where the authors clearly show that MIMO radar is not a cure for all radar problems, but, at the same time, foresee several applications (e.g., high-frequency over-the-horizon radars, ground moving target indication, autonomous vehicles) where it brings great benefits. Bearing in mind that the skepticism against MIMO communications recently turned into widespread adoption and appraisal, there is a chance that the same will happen with MIMO radar in the next decade---if the performance gains, synergies with communication technology, and business cases are sufficiently strong.

\subsection{Vision}

Taken together, the arguments presented above show that the MIMO concept has led to a revolution not only in communications but also in radar, with interesting cross-fertilization of ideas. We foresee that further breakthroughs can be obtained by extending the concept of Massive MIMO to radar. Clearly, by using very large arrays for waveform design, one can radically improve the spatial diversity gain and spatial resolution for target detection, parameter estimation, and interference rejection. However, this is just the beginning of the story. Indeed, the lesson learned from the last decade of research in communications is that Massive MIMO is not merely a system with many antennas but rather a paradigm shift with regards to the modeling, operation, theory, and implementation of MIMO systems. Our vision is that these insights can be applied also to radars, and open a vibrant scientific field, revisit fundamental problems, and possibly enlarge the range of radar applications.

A preliminary work unveiling promising results on \emph{Large-scale MIMO radars} can be found in \cite{Fortunati2019} wherein the authors focus on target detection and show that the high spatial degrees of freedom offered by the array can be exploited to dispense from prior knowledge of the statistical model of the received data. Such knowledge is traditionally seen as crucial for achieving high radar performance since it allows a proper design of the detection algorithm. Unlike MIMO communications, wherein interference is well-modeled as white or colored Gaussian noise, in radar applications this is seldom a realistic hypothesis because of the echoes produced from nearly all surfaces (e.g., ground, sea, and buildings) when illuminated by a radar \cite{Billingsley}. This disturbance is called \emph{clutter} and its statistical characterization is not an easy task since it depends on many factors. Different statistical models are available in the literature \cite{Sangston2012}, which have been used to develop a large variety of detectors. However, experience with real data has exposed their limitations \cite{Fortunati2017}; that is, large differences between the true data and assumed model have been observed. When this happens, the model is said to be mismatched (or misspecified) and the radar performance highly deteriorates. The ultimate objective of any radar system is the development of robust signal processing schemes that guarantee certain performance independently of the clutter distribution. However, this is a challenging open problem due to its complex nature. Possible solutions in the context of Large-scale MIMO radars are developed in \cite{Abla_conf,Abla,Large_scale}, where tools from random matrix theory are used to compute closed-form expressions for the asymptotic false alarm and detection probabilities. These papers rely on the assumption that the detector has a large number of i.i.d.~observations; that is, the target and the functional form of the clutter's statistics maintain constant over a large time interval. This simplification is not true in practice. 
Instead of exploiting random matrix theory tools, the authors in \cite{Fortunati2019} pioneer the combination of misspecification theory \cite{Richmond2015,Fortunati2017} with Large-scale MIMO radars. 
For the practical case of unknown clutter statistics and correlated observations over the array, they build a detector whose asymptotic distribution, as $M\to \infty$, is a $\chi^2$-distribution regardless of the true, but unknown, statistical characterization of the data.

We foresee that the higher degrees of freedom offered by the large-scale MIMO radar technology will enable meeting the continuously growing complexity of military and commercial scenarios. Think, for example, of small \ac{uavs}. The UAV technology has developed rapidly since the beginning of the century and is considered a promising solution for monitoring, transport, safety and disaster management. However, the use of \ac{uavs} in all these domains opens several issues including (among many others) their identification and tracking. Indeed, \ac{uavs} have an RCS of up to three orders-of-magnitude smaller than manned aircrafts \cite{Kurty2017} and tend to operate over shorter distances, slower speeds, lower altitudes and have more variable positions, velocities, and acceleration parameters. All this makes \ac{uavs} very difficult to detect and track with the current radar technology. The extremely increased diversity and spatial resolution of Large-scale MIMO radars may be a viable solution to this issue.

The quest for ever increasing data rates in communications has pushed the carrier frequencies up towards the mmWave bands traditionally occupied by radar. Early studies on the coexistence of communications and radar in these bands took a rather radar-centric approach (see, e.g., \cite{Deng2013b,Maio2015}) wherein the primary concern was to guarantee the detection and estimation performance of the radar by designing waveforms producing a tolerable level of interference on the communication system. The focus has been recently steered back to communication performance \cite{Petropulu2015,Mahal2010,Zheng2018}. We foresee that the high spatial degrees of freedom provided by large antenna arrays in both systems are key to achieving seamless spectral coexistence between radar and communication services \cite{Liu2018}.

\subsection{Open Problems}

It is hard to make a precise list of open problems for a Large-scale MIMO radar since a clear understanding of its advantages and limitations is still missing, and only very preliminary ideas have been developed. Although each radar system has different requirements and constraints that highly depend on its application, the first major problem will be the design of robust signal processing algorithms for detection and estimation, which have a scalable complexity with the number of antennas and are implementable in a centralized or distributed manner depending on the antenna array configuration. 
Ideally, the same antenna arrays can be used for both radar and communications, in the same or adjacent frequency bands. This is illustrated in Fig.~\ref{fig:MIMOradar}, where the radar components have the same configuration as Massive MIMO arrays.

The massive diversity and spatial resolution achieved by Large-scale MIMO radars will inevitably call for new statistical models of the target and clutter. Extensive measurement campaigns will thus be needed to develop such models, which are crucial  to validate the effectiveness of the technology and explore its fundamental limits for detection and estimation. Moreover, testbed development will be highly desired for proof-of-concept and performance evaluation under real-world conditions and to identify further practical challenges that might have been overlooked in the modeling.

Building radar systems with hundreds of antennas will require an economy of scale in manufacturing comparable to what we have seen for short-range sensors. This will not be an easy change since the military industry was, and still is, the dominant user (for surveillance, navigation, and weapon guidance) and developer of radar technology. One option is to use consumer-grade components in radars, instead of military-grade components, but this will introduce hardware distortions that need to be taken into account in the design, analysis, and signal processing.

\vspace{5mm}

\section{Direction 5: Intelligent Massive MIMO}

There is a recent surge of papers applying \emph{Machine Learning} (ML) to various problems in communications \cite{jiang2017machine}. ML can be used to improve the performance of existing algorithms, achieve the performance of close-to-optimal algorithms with reduced implementation complexity, or enable entirely novel use cases which are not feasible with model-based approaches. Some recent examples include MIMO detection \cite{samuel2017deep}, blind channel equalization \cite{caciularu2018blind}, channel estimation \cite{neumann2018learning} and feedback \cite{wen2018deep}, 
Massive MIMO power control \cite{Sanguinetti2018b,Chien2019a},
user positioning \cite{vieira2017deep}, as well as full learning of transceiver implementations \cite{oshea2017introduction}. ML is especially powerful when the system has characteristics that are hard to model or analyze by conventional approaches, but can be learned from data.
Hence, the domain of communications with low-precision analog-to-digital converters (ADCs), or more generally low-cost hardware with impairments, seems a formidable opportunity for ML since non-linear systems are difficult to model and optimal algorithms hard to derive analytically \cite{jeon2017blind, balevi2018one}. One-bit ADCs would enable ultra low-cost Massive MIMO systems with dramatically reduced energy consumption. 

The acquisition of accurate CSI is the possibly most important problem related to Massive MIMO, as highlighted by countless publications on this topic; see \cite{massivemimobook} and reference therein. It is remarkable though that instantaneous CSI is currently only considered valuable during the channel coherence block (e.g., for precoding/receive combining) and is then typically discarded. We argue that storing and analyzing huge quantities of CSI---possibly enriched with side-information, such as user position, CSI on a different frequency band, or transmitter identity---is extremely valuable and might enable significant performance improvements as well as novel use cases. ML is the primary tool to make sense out of the massive CSI data sets that could be easily acquired in each cell. Predicting downlink CSI from uplink CSI \cite{safari2018deep}, uniquely identifying radio transmitters \cite{sankhe2018oracle}, as well as detecting human activity and even emotions \cite{khalili2019wi} (and references therein) are just a few things that can be learned from CSI or, more generally, raw RF samples. We expect many more such applications to arise in the near future. Also, the idea of ``channel charting'' \cite{studer2018channel} (i.e., the positioning of users based on CSI through unsupervised learning on massive data sets) is a very promising future research direction. 

It is typically assumed that antenna arrays are static; that is, the position of individual antennas is fixed and steering of ``beams'' in all directions can be carried out in the digital domain.\footnote{Most existing antenna arrays actually have a physical tilt angle that can be (remotely) controlled to shrink or enlarge the coverage area of the BS.} However, for a fixed array, there are certain things that cannot be done in the digital domain. For example, energy cannot be ``beamed'' around a large object that blocks a crucial propagation path. To deal with such problems, an interesting avenue for future research is the study of arrays with flexible geometry that could optimally position themselves or their elements with some degrees of freedom. For instance, the array aperture could be dynamically optimized to provide high spatial resolution to crowded areas while possibly reducing resolution in less crowded parts of a cell. 
A key benefit compared to ELAAs and Holographic Massive MIMO is that less hardware is needed if the geometry can be adapted to the current user population.
To bring this idea even one step further, one could not only dynamically optimize the array geometry but also the radio environment itself, for example,  through software-controlled metasurfaces \cite{Subrt2012a,liaskos2018new} whose RF reflection properties can be dynamically adapted to positively influence or optimize the radio environment {(see Direction 2)}. ML can be used to optimally control such ``smart radio environments'' \cite{zappone2019wireless} together with the array position and geometry.

\subsection{Vision}

In our vision, ML will become a first-class citizen in future communication systems and enable truly \emph{Intelligent Massive MIMO}. ML will be used whenever a good model is lacking, a model is available but it is intractable for analysis, or when model-based algorithms are too complex for practical implementation. The question is not if but rather when and where we will see the first ML solutions appear in products. Massive MIMO systems will store huge quantities of CSI in each cell (e.g., RF samples) and ML will be used to exploit this novel resource for various purposes.

\begin{figure*}[t!]
	\centering \vspace{-2mm}
	\begin{overpic}[width=1.5\columnwidth,tics=10]{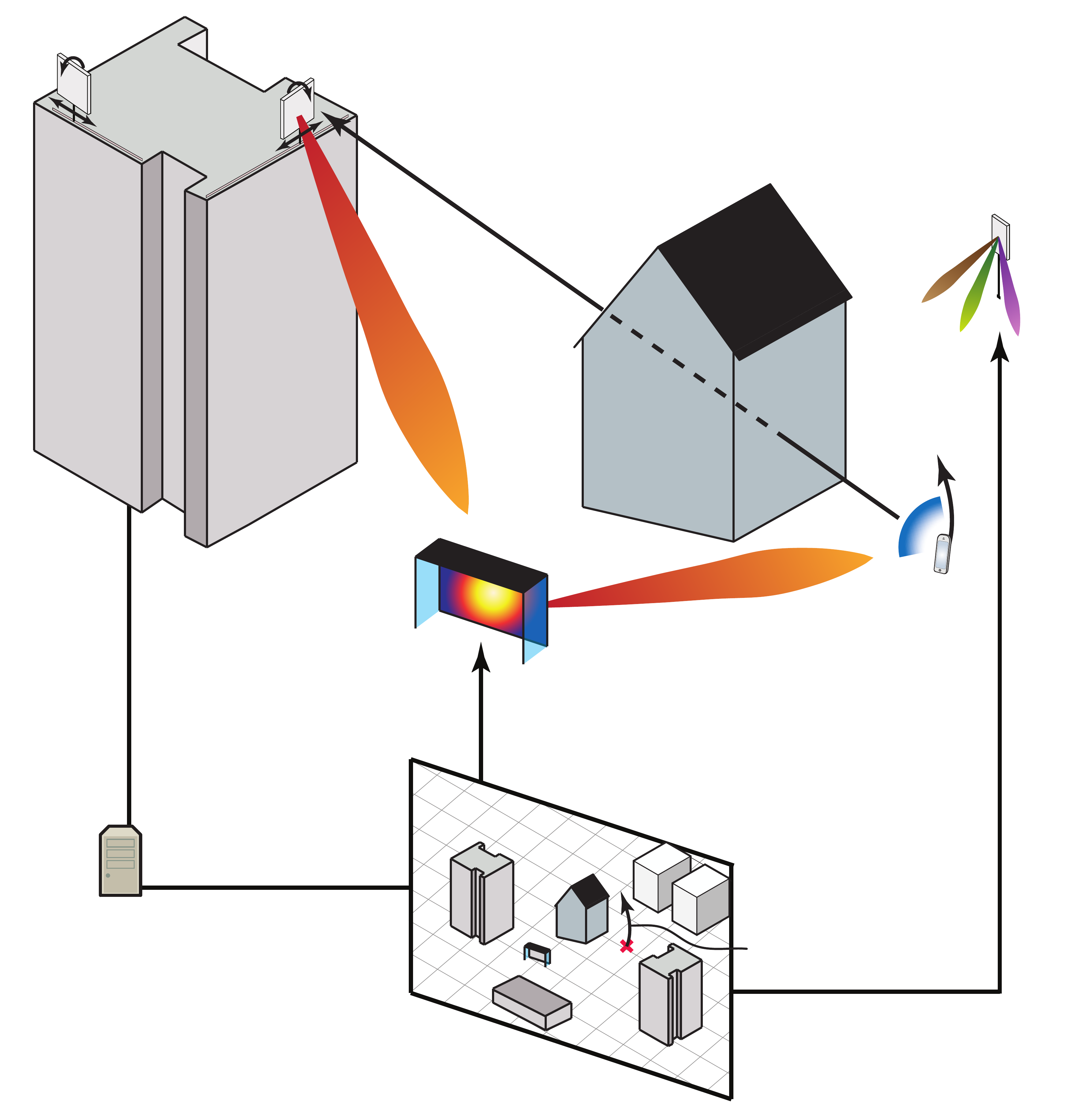}
		 \put (-15,98) {(a) Optimal array placement}
		 \put (-15,95) {and configuration}
	  \put (29,92) {(b) ML-based channel estimation}
	  \put (14,40) {(c) Configure metasurface}
	  \put (14,37) {to reflect signal favorably}
	  \put (14,34) {towards the user}
	  \put (40,83) {Uplink pilot}
	  \put (36,4) {Channel chart}
	  \put (36,1) {with position}
	  \put (91,65) {(d) Predicted trajectory}
	  \put (91,62) {anticipates handover}
	  \put (91,59) {and target BS begins}
	  \put (91,56) {optimized beam}
	  \put (91,53) {alignment}
	  \put (67,14.5) {Predicted trajectory}
	   \put (0,16) {(e) Baseband processor}
	   \put (0,13) {augmented with ML}
\end{overpic} 

	\caption{ML can be used to refine almost any part of a conventional Massive MIMO system, turning it into an Intelligent Massive MIMO system. This figure exemplifies five different parts of the network where ML can be applied.}
	\label{fig:ml-example}
\end{figure*}
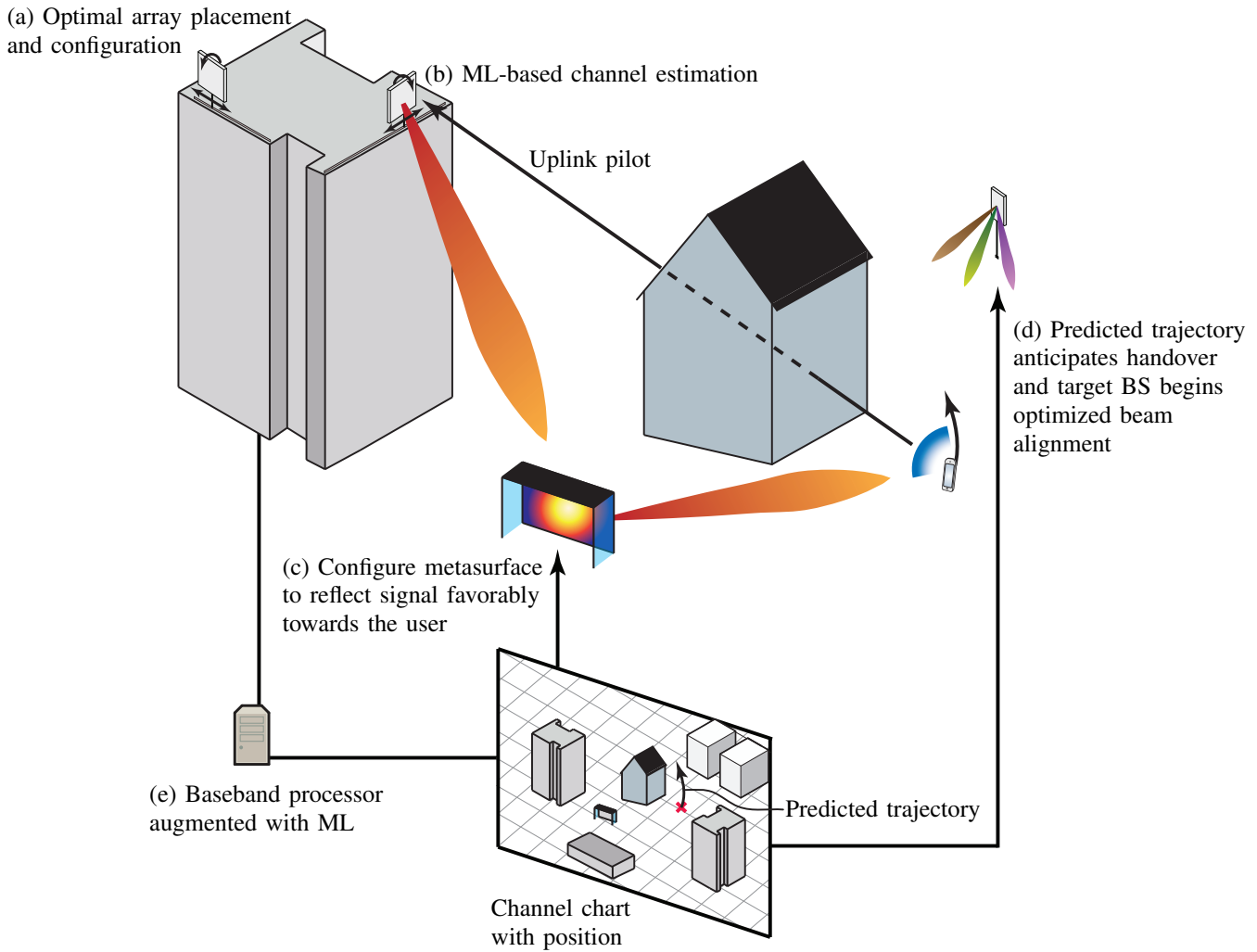

Fig.~\ref{fig:ml-example} shows an example of how multiple ML-based use cases can be combined: (a) Multiple connected antenna arrays can be positioned on rails on the roof of a building, and also the orientation and position of their individual antennas can be controlled. \emph{Reinforcement Learning} (RL) is used to optimize the array configuration over time. (b) A user's channel is estimated through an uplink pilot. The channel estimator itself has learned to exploit the spatial characteristics of the local radio environment. Next, channel charting or an explicit form of user positioning (as described earlier) is used to locate the user within a three-dimensional map. The large number of antennas makes CSI-based positioning robust with respect to changes in the environment, for example, due to moving reflectors/scatterers. (c) The position of a user (or its position within the channel chart) is used to configure one or multiple metasurfaces such that the radio waves are reflected favorably towards the user, which in this example is located behind a building that blocks the \ac{los} path from the BS. (d) Deep learning is used to identify the user based on RF fingerprinting, and its speed and trajectory are estimated. Depending on the outcome, neighboring cells are informed about possible handovers. In the figure, a mmWave BS is preconfigured to use a certain set of beams to speed-up the beam alignment procedure with the user. (e) We expect that ML will also play a major role in the digital signal processing carried out in the baseband processor of the BS. For example, multi-user detection with low-precision ADCs as well as digital predistortion for the BS's power amplifiers could be enhanced by deep neural networks.

Fig.~\ref{fig:ml-example} barely captures the breadth of possible applications but hopefully convinces the reader that ML is going to impact almost every part of future communication systems. That is why call this Intelligent Massive MIMO.

\subsection{Open Problems}

Before ML can be successfully used in communication systems, several major obstacles need to be overcome. Since each application comes with its own requirements and constraints, we will only mention a few universal problems here.

To some extent, training data can be seen as the ``source code'' of ML systems. This implies that---similar to source code which needs to be debugged before compilation---data must be cleaned and biases, outliers, as well as anomalies removed before training. Today, acquiring data from running networks is very difficult at best and hardly any open data sets are available for research. The latter has been the main enabler for breakthroughs in other domains such as natural language processing or computer vision, where ImageNet  \cite{deng2009imagenet} is the standard database used for object recognition.  The cost of data acquisition, especially for the physical layer that generates huge quantities of data, is not negligible and the data volume often surpasses the user data that is actually transmitted. Thus, ML applications will only make it into products if their benefits outweigh the related cost. Hence, identifying the right use cases for ML in communications is the first priority, while the tweaking of neural networks and learning procedures for maximum performance has a lower priority.

Communication systems require ML inference with hard real-time constraints at the micro- to nanosecond timescales. This is one to two orders of magnitude faster than what is required by the most demanding computer vision applications (e.g., self-driving cars). Due to this reason and the huge data quantities involved, the inference must be carried out close to where the digital baseband processing resides; that is, in a BS or user device. Since processing power is limited on such devices, neural networks that are typically trained on GPUs or other specialized hardware in data centers must be compressed and quantized to be efficiently implemented in radio hardware \cite{carreira2017model, morphnet2018gordon}, for example, using FPGAs and ASICs. A related aspect is that one cannot expect to train a model once and then reuse it forever. In many instances, fast online training on the device is needed. However, gradient-based approaches might be too slow and alternatives must be found. The above challenges are crucial for and specific to communications and must be addressed by our community since ML research in other domains has not yet addressed them sufficiently. 

At first glance, ML and, particularly, RL seem to have tremendous potential to deal with the increasingly complex optimization problems that arise in radio resource management of large-scale cellular networks \cite{calabrese2018learning}. It seems plausible that even simple ML algorithms could outperform the man-made heuristics that are currently implemented in commercial products. However, it is challenging to put RL algorithms into practice because learning on real systems tends to be slow (and would possibly not even be allowed by operators) while today's simulation environments are unlikely to be sufficiently representative of the real world. Thus, unless we develop sufficiently realistic simulation environments for large communication networks (e.g., for OpenAI Gym \cite{openaigym}), ideally based on ray tracing, it seems unlikely that RL-based solutions will be employed in practice in the foreseeable future.

An often neglected aspect concerning the use of ML in complex systems is the \emph{technical debt} \cite{hidden2015sculley} and the ML model management challenges \cite{challenges2018schelter} it creates. For example, if one ML model depends on the outputs of another, or if the outputs of multiple models are mixed together as inputs to another model, complicated data dependencies arise and seemingly disjointed systems might be unintentionally coupled. Moreover, when one model or the input data to one of the models change, everything must be retrained from scratch. Another potential problem is the creation of undesired feedback loops where an ML model determines its own future training data or two systems influence each other through decisions taken in a shared environment. Although we are still at the infancy of ML applications in communications, it is important to keep these issues in mind which may quickly gain in relevance once ML solutions find widespread adoption.

\section{Conclusions}

This paper has outlined five promising research directions that involve Massive MIMO, digital beamforming, and/or antenna arrays. We hope that this will inspire the large number of researchers that are currently active in the Massive MIMO and mmWave communication fields to find new paths and open problems to tackle in the coming decade. The directions need not be considered independently, but there are many synergies and opportunities to combine them. Admittedly, none of the five directions are entirely new but we are expecting them to have a renaissance now when the Massive MIMO technology becomes mainstream and omnipresent.
The future will tell how quickly, and to what extent, our visions for the five directions will materialize, but it might go faster than we can imagine today. Just bear in mind that Massive MIMO was generally considered science fiction ten years ago, and now it is a reality.

We have not provided an exhaustive list of prospective research directions that involve antenna arrays. In fact, there are more disruptive, futuristic, and untouched paths to follow for those who dare. For example, what would happen if we combine the Massive MIMO concept with quantum communications, signaling at frequencies above the radio  band (i.e., $>300$\,GHz), or molecular communications?

\bibliographystyle{IEEEtran}
\bibliography{IEEEabrv,refs}

\end{document}